\documentclass[nofootinbib,10pt,a4paper,twocolumn]{revtex4-1}

\usepackage{amsmath}
\usepackage{amsfonts}
\usepackage{amssymb}
\usepackage{amsthm}
\usepackage{mathrsfs}
\usepackage{dsfont}
\usepackage{esint}
\usepackage{graphicx}

\DeclareMathAlphabet{\mathpzc}{OT1}{pzc}{m}{it}

	\newcommand{\PropTo}{\propto}
	\newcommand{\AsymEq}{\sim}
	\newcommand{\ApproxEq}{\approx}

	\newcommand{\Ket}[1]{\left|#1\right\rangle}
	\newcommand{\sKet}[1]{|#1\rangle}
	\newcommand{\Bra}[1]{\left\langle#1\right|}
	\newcommand{\sBra}[1]{\langle#1|}
	\newcommand{\BK}[2]{\left\langle#1|#2\right\rangle}
	\newcommand{\sBK}[2]{\langle#1|#2\rangle}
	\newcommand{\BAK}[3]{\Bra{#1}#2\Ket{#3}} 
	\newcommand{\sBAK}[3]{\sBra{#1}#2\sKet{#3}} 
	\newcommand{\KB}[2]{\left|#1\right\rangle\!\left\langle #2\right|}

	\newcommand{\V}[1]{\ensuremath{\boldsymbol{#1}}}			

	\newcommand{\mr}[1]{\mathrm{#1}}			

	\newcommand{\br}[1]{\left( #1 \right)}
	\newcommand{\brr}[1]{\left[ #1 \right]}
	
	\newcommand{\of}[1]{\!\br{#1}}
	\newcommand{\off}[1]{\!\brr{#1}}
	
	\newcommand{\sbr}[1]{( #1 )}
	\newcommand{\sbrr}[1]{[ #1 ]}
	\newcommand{\sbrrr}[1]{\{ #1 \}}
	\newcommand{\sof}[1]{\!\sbr{#1}}
	\newcommand{\soff}[1]{\!\sbrr{#1}}
	\newcommand{\sofff}[1]{\!\sbrrr{#1}}

	\newcommand{\Sum}[2]{\sum\limits_{#1}^{#2}}
	\newcommand{\Prod}[2]{\prod\limits_{#1}^{#2}}
	
	\newcommand{\Int}[3]{\int\limits_{#1}^{#2}\mr{d}#3\,}
	
	\newcommand{\sSum}[2]{\sum_{#1}^{#2}}
	\newcommand{\sProd}[2]{\prod_{#1}^{#2}}
	
	\newcommand{\sInt}[3]{\int_{#1}^{#2}\mr{d}#3\,}

	\newcommand{\EA}[1]{\xpc{#1}}

	\newcommand{\xpc}[1]{\left\langle #1 \right\rangle}

	\newcommand{\sEA}[1]{\sxpc{#1}}

	\newcommand{\sxpc}[1]{\langle #1 \rangle}

	\newcommand{\Reals}{\ensuremath{\mathbb{R}} }

	\newcommand{\Integers}{\ensuremath{\mathbb{Z}} }

	\newcommand{\D}{\mathrm{d}}

	\newcommand{\sLandau}[1]{\mathpzc{O}\sof{#1}}

		\newcommand{\Max}[2]{\max\of{#1,#2}}
		\newcommand{\Abs}[1]{\left\vert #1 \right\vert}
		\newcommand{\sAbs}[1]{\vert #1 \vert}
		
		\newcommand{\sNorm}[1]{\| #1 \|}

		\newcommand{\Sign}[1]{\mr{sgn}\of{#1}}
		\newcommand{\Id}{\mathds{1}}

		\newcommand{\Trace}[1]{\mathrm{Tr}\off{#1}}
		\newcommand{\sTrace}[1]{\mathrm{Tr}\soff{#1}}

		\newcommand{\Gma}[1]{\Gamma\of{#1}}
		\newcommand{\sGma}[1]{\Gamma\sof{#1}}

		\newcommand{\BesselJ}[2]{J_{#1}\of{#2}}
		\newcommand{\sBesselJ}[2]{J_{#1}\sof{#2}}

\newcommand{\Brillouin}{\ensuremath{\mathbb{B}}}
\newcommand{\Hilbert}[1]{\ensuremath{\br{\mathcal{H} #1}}}
\newcommand{\sHilbert}[1]{\ensuremath{\sbr{\mathcal{H} #1}}}

\newcommand{\Resolv}{\ensuremath{\hat{\mathcal{U}}}}
\newcommand{\Detect}{\ensuremath{\hat{D}}}
\newcommand{\Ham}{\ensuremath{\hat{H}}}
\newcommand{\TEO}{\ensuremath{\hat{U}}}

\newcommand{\PsiIn}{\ensuremath{\psi_\text{in}}}
\newcommand{\PsiDet}{\ensuremath{\psi_\text{d}}}

\newcommand{\Fourier}[1]{\mathcal{F}\sofff{#1}}

\begin{document}
	\title{The spectral dimension controls the decay of the quantum first detection probability}
	\author{Felix Thiel}
	\email{thiel@posteo.de}
	\affiliation{Bar-Ilan University, Department of Physics, 5290002 Ramat Gan, Israel}

	\author{David A. Kessler}
	\affiliation{Bar-Ilan University, Department of Physics, 5290002 Ramat Gan, Israel}

	\author{Eli Barkai}
	\affiliation{Bar-Ilan University, Department of Physics, 5290002 Ramat Gan, Israel}
	\begin{abstract}
		We consider a quantum system that is initially localized at $\V{x}_\text{in}$ and that is repeatedly projectively probed with a fixed period $\tau$ at position $\V{x}_\text{d}$.
		We ask for the probability that the system is detected in $\V{x}_\text{d}$ for the very first time, $F_n$, where $n$ is the number of detection attempts.
		We relate the asymptotic decay and oscillations of $F_n$ with the system's energy spectrum, which is assumed to be absolutely continuous.
		In particular $F_n$ is determined by the Hamiltonian's measurement spectral density of states (MSDOS) $f\sof{E}$ that is closely related to the density of energy states (DOS).
		We find that $F_n$ decays like a power law whose exponent is determined by the power law exponent $d_S$ of $f\sof{E}$ around its singularities $E^*$.
		Our findings are analogous to the classical first passage theory of random walks.
		In contrast to the classical case, the decay of $F_n$ is accompanied by oscillations with frequencies that are determined by the singularities $E^*$.
		This gives rise to critical detection periods $\tau_c$ at which the oscillations disappear.
		In the ordinary case $d_S$ can be identified with the spectral dimension found in the DOS.
		Furthermore, the singularities $E^*$ are the van Hove singularities of the DOS in this case.
		We find that the asymptotic statistics of $F_n$ depend crucially on the initial and detection state and can be wildly different for out-of-the-ordinary states, which is in sharp contrast to the classical theory.
		The properties of the first detection probabilities can alternatively be derived from the transition amplitudes.
		All our results are confirmed by numerical simulations of the tight-binding model, and of a free particle in continuous space both with a normal and with an anomalous dispersion relation.
		We provide explicit asymptotic formulae for the first detection probability in these models.
	\end{abstract}

	\date{Manuscript of \today}

	\maketitle

	\section{Introduction and main results}
		One of the first questions addressed in elementary physics is how much time does it take an object to reach its destination.
		In non-equilibrium statistical physics this is the first-arrival problem which is fundamental for diffusion controlled chemical reactions and other search problems \cite{Redner2007-0,Benichou2005-0,Benichou2011-0,Godec2016-0,Godec2016-1}.
		Seemingly similar to the motion of diffusing particles, quantum systems appear to evolve randomly.
		One might therefore ask for the probability that a quantum system initially localized at $\V{x}_\text{in}$ arrives at a target position $\V{x}_\text{d}$ at time $t$ for the first time.
		In a more general setup, the target $\V{x}_\text{d}$ and initial position $\V{x}_\text{in}$ can be replaced with any valid states from the Hilbert space, $\sKet{\PsiDet}$ or $\sKet{\PsiIn}$, respectively.
		This question leads to complications.
		Soon after the establishment of quantum theory, it became clear that there can be no self-adjoint operator that represents the arrival time \cite{Allcock1969-0}.
		Still, since the concept of arrival times is so very basic and intuitive, efforts have been made to define and then calculate the arrival time by a plethora of other methods: 
		imposing special boundary conditions on the Schr\"odinger equation \cite{Kumar1985-0}, introducing stochastic forces \cite{Lumpkin1995-0}, or via imaginary potentials \cite{Krapivsky2014-0}.
		It was also suggested to define the arrival time in terms of positive-operator-valued measures \cite{Kijowski1974-0,Sombillo2016-0}.
		In the operative approach, one tries to incorporate the detector directly into the model \cite{Aharonov1998-0,Damborenea2002-0}.
		Moreover, some authors used the concept of decoherent histories to describe the path of the system until it reaches the target state \cite{Anastopoulos2006-0,Halliwell2009-0,Halliwell2009-1}.

		One of the major conceptual problems is that a quantum particle does not possess a trajectory -- it can not be tracked in the classical sense.
		To infer the position or state of a quantum system, the observer must perform a measurement that will collapse the system's wave function.
		The correct moment to perform the measurement to achieve success is of course unknown a priori and too frequent measurement will lock the system's dynamics via the quantum Zeno effect \cite{Misra1977-0,Itano1990-0}.
		A very pragmatic solution to the dilemma is the introduction of a detection protocol \cite{Gruenbaum2013-0,Dhar2015-0}.
		Here, the observer decides before the experiment when he will attempt detection.
		This allows the theoretician to weave the backfire of the initial unsuccessful measurements into the remaining (unitary) dynamics of the quantum system.
		Periodically measured systems have also been considered in different contexts, e.g. using the measurements as a heat bath \cite{Yi2011-0}.

		A popular choice is to attempt detection at fixed intervals of duration $\tau$, the so-called stroboscopic detection protocol.
		Such a setup was considered in \cite{Krovi2006-0,Gruenbaum2013-0,Montero2013-0,Bourgain2014-0,Dhar2015-0,Dhar2015-1,Sinkovicz2015-0,Sinkovicz2016-0,Lahiri2017-0} and by the authors in \cite{Friedman2017-0,Friedman2017-1,Thiel2018-0}.
		This is also the approach of the present work.
		The detection protocol shifts the emphasis of the question: One does not ask for the first arrival of the system, but rather for its {\em first detection} in the target state.
		Particularly, we ask what is the probability $F_n$ that the system is first detected in the target at the $n$-th attempt.

		The first detection problem is particularly relevant from the perspective of quantum computing.
		It is closely related to the quantum search problem \cite{Grover1997-0,Ambainis2001-0,Aaronson2003-0,Childs2004-0,Li2017-0} and translates to the question of when a computation result becomes available.
		The popular search algorithms of Refs.~\cite{Grover1997-0,Childs2004-0} focus on tuning a quantum system such that it most effectively transforms some fixed initial state into some a priori unknown oracle state.
		Our approach is different in that we fix a target state and ask for the time when it is reached.
		Hence, our focus is on the investigation of $F_n$, which can later be optimized.

		The canonical tight-binding model will serve as an example throughout the discussion.
		It is most conveniently realized in wave-guide-lattice experiments \cite{Perets2008-0}, which could easily be modified to our setup.
		The first detection of a quantum walker in the one-dimensional tight-binding model was discussed in Refs.~\cite{Dhar2015-0,Dhar2015-1,Friedman2017-0,Friedman2017-1,Thiel2018-0}.
		It was found that for large times the first detection probability decays like a power law with exponent $-3$ upon which strong oscillations are superimposed.
		Two dimensional systems have also been considered numerically and within a perturbation approach where different exponents were reported \cite{Dhar2015-1}.
		In a more general setting, e.g. in higher dimensions including fractal systems, it is still an open question what controls the large $n$ behavior of $F_n$.
		In this paper, we will focus on systems with a continuous energy spectrum, which is shown to give rise to the power-law decay of $F_n$.

		In the classical theory of random walks, the first passage probability $F_n^\text{(cl)}$ also decays as a power law, with an exponent that depends on the {\em spectral dimension}, sometimes also called fracton, or harmonic, dimension.
		This important quantity first appeared in the discussion of transport in fractal systems \cite{Alexander1982-0,Hughes1995-0}, where it was found that many traditionally equivalent definitions of dimensionality surprisingly give non-integer values and do not coincide.
		The spectral dimension is defined as the power law exponent found in the density of energy states (DOS) $\rho\sof{E}$, which behaves like $E^{d_S^\text{DOS}/2-1}$ for small energies $E>0$.
		(Throughout our work, we shift the minimal possible energy to $E=0$.)
		The DOS is a property of the Hamiltonian and, due to the ubiquity of the Laplacian, also appears in countless other physical problems, e.g. lattice vibrations \cite{Montroll1947-0,vanHove1953-0,Alexander1981-0,Hughes1995-0}.
		In non-fractal systems with Euclidean dimension $d$, one finds $d_S^\text{DOS}=d$; the Sierpiensky gasket -- a popular fractal -- has $d_S^\text{DOS} = \ln 9 / \ln 5 \ApproxEq 1.365$ \cite{Alexander1982-0}.
		For a classical random walk, the probability $P_n^\text{(cl)}\sof{\V{x}|\V{x}}$ to return to its initial position at the $n$-th step can be expressed as the Laplace transform of the DOS \cite{Alexander1981-0}.
		Consequently this quantity decays as a power law with exponent $-d_S^\text{DOS}/2$.
		The first passage probability is then computed from the return probability \cite{Redner2007-0}, and also decays as a power law with exponent $-\Max{d_S^\text{DOS}/2}{2-d_S^\text{DOS}/2}$.
		Logarithmic corrections to the power law appear in the critical dimension $d_S^\text{DOS} = 2$.
		The DOS, the first passage probability and the return probability $P_n^\text{(cl)}\sof{\V{x}|\V{x}}$ are in a triangle relation to each other.

		\begin{figure}
			\includegraphics[width=0.5\columnwidth]{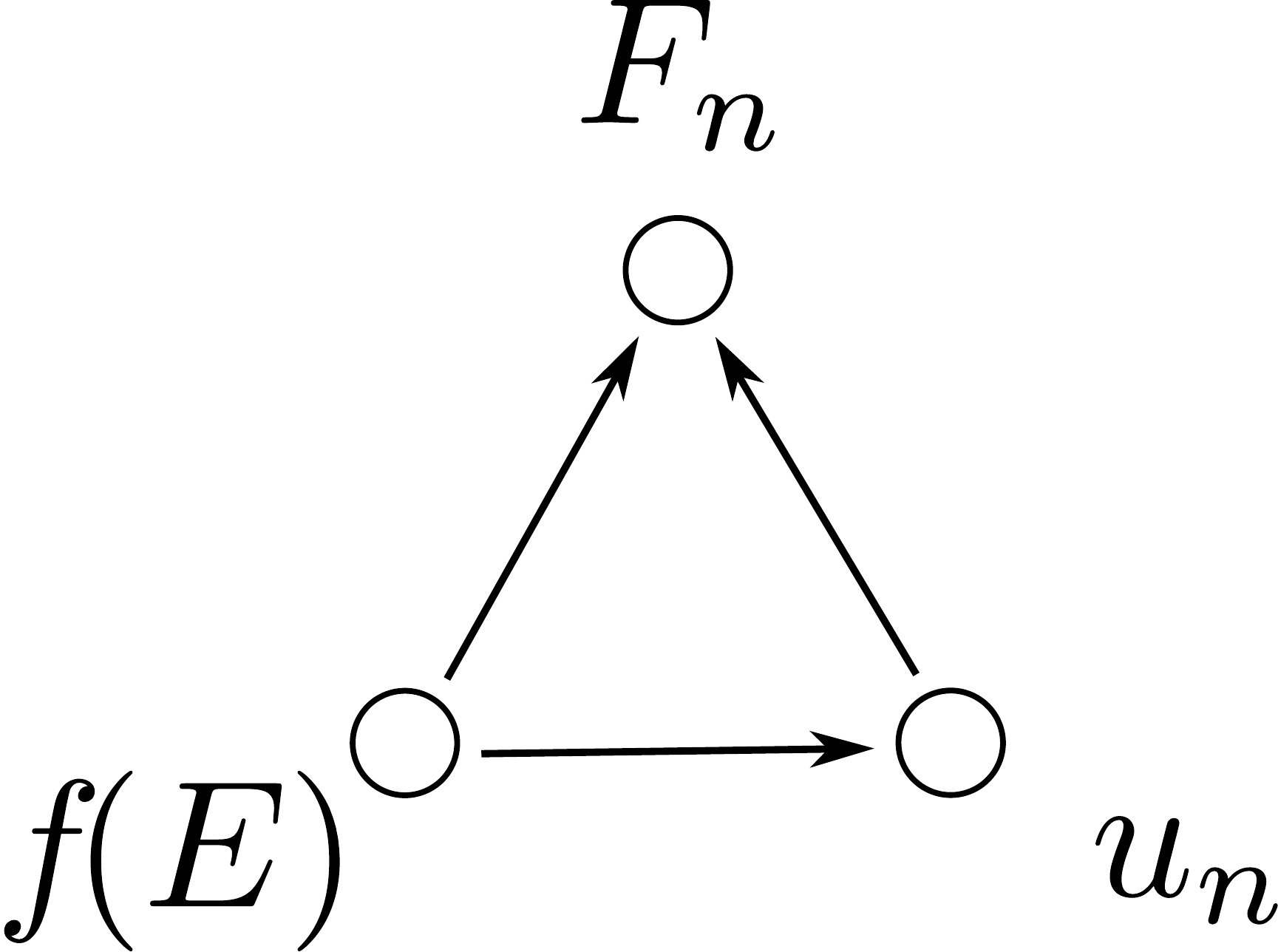}
			\caption{
				A sketch symbolizing the relations between the energy spectrum (represented by the measurement spectral density of states $f\sof{E}$, see section~\ref{sec:SpecMeas}), the return amplitudes $u_n$ [see Eq.~\eqref{eq:DefRA}], and the first detection probabilities $F_n$.
				We present two ways to calculate $F_n$: Directly from the spectrum or via the return amplitudes.
				\label{fig:Triangle}
			}
		\end{figure}
		The very same program is carried out in this article in the quantum case.
		We relate the Hamiltonian's spectral properties to the first detection probability $F_n$ as well as to the {\em amplitude} of return.
		A subtle difference between classical first passage theory and the quantum first detection problem is that the DOS is not the relevant quantity, but rather the so-called {\em measurement spectral density of states} (MSDOS) $f\sof{E}$ of the Hamiltonian (defined below).
		This key quantity is well known in the mathematical literature \cite{Marchetti2012-0}.
		It is closely related -- and sometimes equal -- to the DOS $\rho\sof{E}$.
		The difference is that the MSDOS not only summarizes the properties of the Hamiltonian, but {\em also those of the initial and detection states.}
		In layman's terms, it combines information about available energy states with information about the initial or detection states' overlap with these energies.
		Hence it allows one to concentrate only on the relevant components of the energy spectrum.

		From the power-law behavior $\sAbs{E-E^*}^{d_S/2-1}$ of $f\sof{E}$ around its singularities $E^*$, a spectral dimension $d_S$ can be defined.
		In this article we discriminate between two classes of quantum states.
		For ``ordinary'' states, $f\sof{E}$ and $\rho\sof{E}$ share the positions and exponents of their singularities, that means $E^*$ can be identified with a van Hove singularity and $d_S = d_S^\text{DOS}$.
		However, there are out-of-the-ordinary states for which this identification is not possible; $d_S^\text{DOS}$ and $d_S$ may assume different values.
		It is the latter exponent that determines the behavior of the system's transition amplitudes \cite{Marchetti2012-0} -- which decay as $\text{time}^{-d_S/2}$ -- as well as the first-detection properties.
		Just as in the classical case, the MSDOS $f\sof{E}$, the first detection probabilities $F_n$, and the (later precisely defined) return amplitudes $u_n$, are cast into a triangle relationship; see Fig.~\ref{fig:Triangle} for an illustration.
		The triangle enables us to compute $F_n$ from the MSDOS $f\sof{E}$, or alternatively from the return amplitudes $u_n$, depending on analytical convenience or on the theoretician's taste.

		Invoking the MSDOS, we show that the power-law decay of $F_n$ is generic for systems with continuous energy spectrum and that its exponent depends only on the spectral dimension $d_S$ found in $f\sof{E}$.
		Our main result is
		\begin{equation}
			F_n
			\AsymEq
			\Abs{
				\Sum{l=0}{L'-1}
				F_{l,d_S} e^{-in\frac{\tau E^*_l}{\hbar}}
			}^2 
			\times
			\left\{ \begin{aligned}
					\frac{1}{n^{4 - d_S}}, & \; d_S < 2 \\
					\frac{1}{n^2 \ln^4 n}, & \; d_S = 2 \\
					\frac{1}{n^{d_S}}, & \; d_S > 2 \\
			\end{aligned} \right.
			.
		\label{eq:AsymFDP}
		\end{equation}
		The quantum power law exponent is exactly double the classical exponent \cite{Redner2007-0}.
		This can be understood in a hand-waving fashion by invoking the fact that one deals with amplitudes in the quantum problem.
		Although both theories are developed along the same lines, the final squaring operation in going from amplitudes to probabilities doubles the resulting exponents.
		Furthermore, we again find the critical dimension to be $d_S = 2$.
		Beside the power law decay, oscillations are typically found and they are described by the $\sAbs{\cdots}^2$ term.
		These oscillations do not occur in every system, because their frequency can be tuned with the detection period $\tau$ and because it depends on the number $L'$ of non-analytic points $E^*_l$ of the MSDOS.
		The oscillations, manifest in the complex exponentials in Eq.~\eqref{eq:AsymFDP}, are a surprising addition from the classical point of view.
		From the quantum perspective, they are easily understood as interference phenomena.

		In the ``ordinary'' case, the spectral dimension $d_S$ found in $f\sof{E}$ is equal to $d_S^\text{DOS}$, the spectral dimension found in the DOS.
		Furthermore, the singularities $E^*_l$ of $f\sof{E}$ can be identified with the {\em van Hove singularities} of the DOS \cite{vanHove1953-0}.
		Since $d_S^\text{DOS}$, as well as the van Hove singularities are properties of the DOS and thus of the Hamiltonian, our result, Eq.~\eqref{eq:AsymFDP}, is ``robust'', in the sense, that a different choice of initial and detection state only changes the amplitudes $F_{l,d_S}$, but neither the power law nor the frequencies of the oscillations.
		In the classical theory, this is where the story ends, but our quantum problem features an epilogue.
		Due to the possibility of superposition of quantum states, the MSDOS $f\sof{E}$ can be wildly different from the DOS $\rho\sof{E}$.
		This can go so far that the singularities $E^*_l$ can not be identified with the van Hove singularities or that even the spectral dimensions do not coincide, i.e. $d_S \ne d_S^\text{DOS}$.
		Consequently, in the quantum first detection probability may depend sensitively on the particular choice of initial and detection state, which is a considerable departure from the classical point of view.
		Eq.~\eqref{eq:AsymFDP} is still valid in this out-of-the-ordinary, but the involved quantities can not be inferred from the DOS anymore.

		Throughout the article we will demonstrate our reasoning in the tight-binding model in $d$ dimensions with Hamiltonian:
		\begin{equation}
			\Ham
			:=
			-\gamma
			\Sum{\V{x}\in\Omega}{} \Ket{\V{x}} 
			\Sum{j=1}{d} \brr{
				\Bra{\V{x}+a\V{e}_j}
				+
				\Bra{\V{x}-a\V{e}_j}
				- 2\Bra{\V{x}}
			}
			,
		\label{eq:TBHamiltonian}
		\end{equation}
		where the energy constant $\gamma$ determines the strength of the nearest neighbor hopping, $\sKet{\V{x}}$ is a position (lattice-site) eigenstate and $\V{e}_j$ is the unit vector in the $j$-th coordinate.
		This model describes a particle moving coherently on an infinite simple cubic lattice with lattice constant $a$.
		We stress though that our result, Eq.~\eqref{eq:AsymFDP}, is fairly generic and holds for any system with a continuous energy spectrum, although with different amplitudes $F_{l,d_S}$.
		(Of course, as we briefly mentioned, terms and conditions apply.
		These are made explicit later in the text.)
		To strengthen this claim, we will also consider a free particle in continuous space.
		Furthermore, Eq.~\eqref{eq:AsymFDP} is also applicable for {\em fractional} spectral dimensions, and we demonstrate this in a free particle with an anomalous dispersion relation.
		
		The rest of the paper is organized as follows:
		In section \ref{sec:Strobo} we explain the stroboscopic detection protocol, i.e. how the detection process is added to the system's natural dynamics.
		We review how to formally obtain the first detection probability using generating functions \cite{Friedman2017-0}.
		Then in section \ref{sec:SpecMeas} we present the main conceptual tool in our investigations, the MSDOS, which is closely related to the DOS.
		Using these, we show that the first detection probability can be represented as a Fourier transform.
		Section \ref{sec:LargeN} follows with an asymptotic formula for Fourier transforms used to derive Eq.~\eqref{eq:AsymFDP}.
		The same formula can be applied to the system's free evolution unperturbed by measurement.
		This opens up an alternative way to compute $F_n$, which is also done in this section.
		Throughout these derivations, we illustrate our reasoning using the tight-binding model.
		In the last two sections, \ref{sec:FreePart} and \ref{sec:Levy}, we compute the first detection probability for two other example models: the free particle in continuous space and a L\'evy particle.
		We close the article with discussion in section \ref{sec:Disc}.
		Derivations that interrupt the flow of presentation have been relegated to the appendices.
		Appendices \ref{app:Plemelj} and \ref{app:Tauber} are concerned with an analogue to the Sokhotski-Plemelj formula on the unit circle and an asymptotic formula for Fourier transforms, respectively.
		In Appendix \ref{app:Arrival}, we discuss the problem of first detected arrival.
		Finally, Appendix \ref{app:Even} treats the case of even dimensions, where logarithmic corrections appear.

	\section{The stroboscopic detection protocol}
	\label{sec:Strobo}
		\begin{figure}
			\includegraphics[width=0.99\columnwidth]{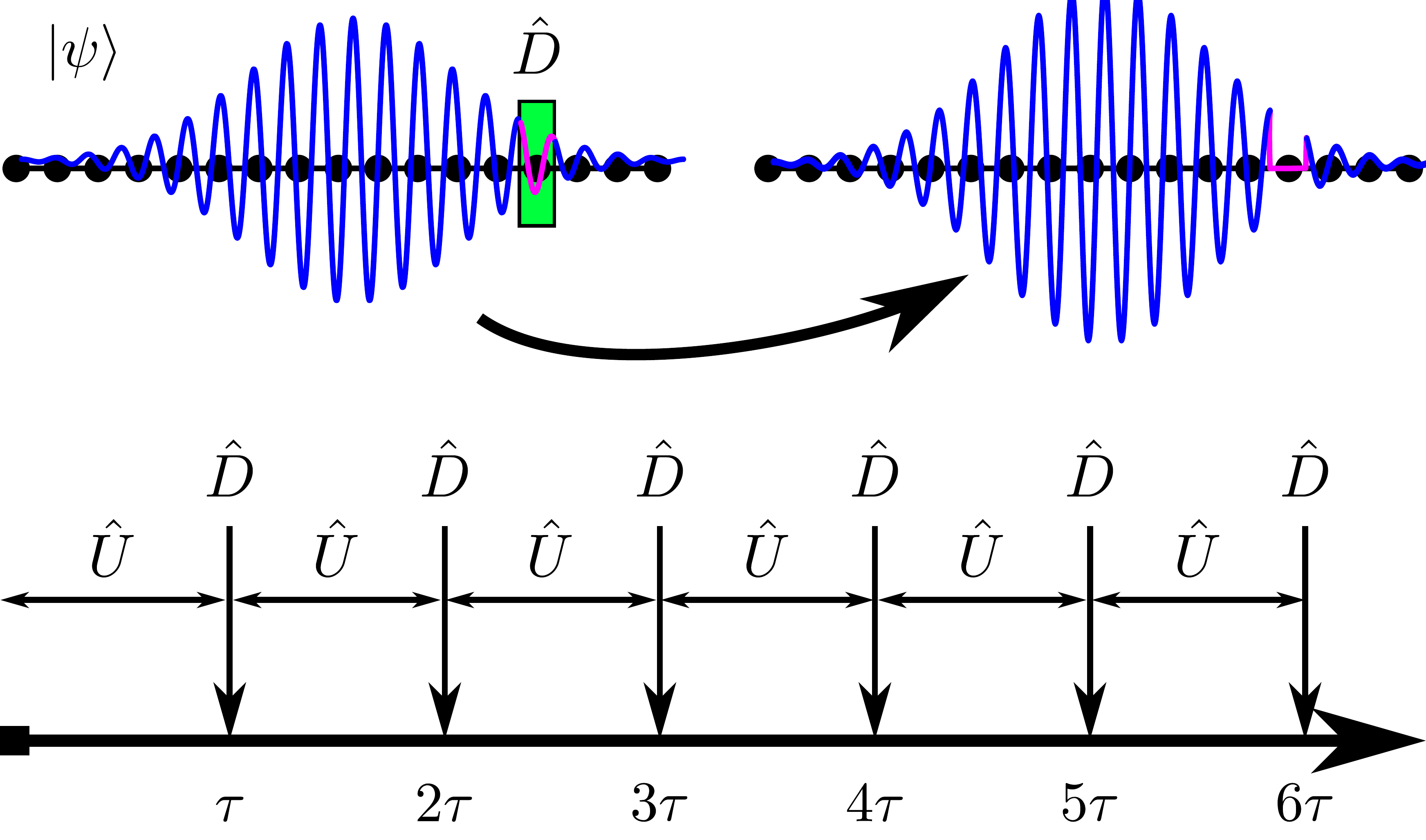}
			\caption{
				Sketch of the detection protocol.
				The system evolves unitarily for $\tau$ time units with the evolution operator $\TEO$ in between the detection attempts that are performed with the projection operator $\Detect$.
				The detection is a strong measurement that erases the wave function in the target site, when it was unsuccessful.
				\label{fig:DetProt}
			}
		\end{figure}
		In this section, we will review the derivation of the first detection probabilities using the renewal equation approach \cite{Friedman2017-0,Friedman2017-1}.
		This sets the ground for a reformulation in terms of the system's energy spectrum, which is found in the next section.

		The system is initially prepared in the state $\sKet{\PsiIn}$.
		In the tight-binding model of Eq.~\eqref{eq:TBHamiltonian}, we can identify $\sKet{\PsiIn} = \sKet{\V{x}_\text{in}}$ with a position eigenstate on the lattice, but in continuous-space systems this must be avoided due to the uncertainty principle.
		In a general setup $\sKet{\PsiIn}$ could be any state from the Hilbert space of the system.
		We will consider both situations (lattice and continuous-space) in the examples.

		The main idea to the first detection problem is to fix the times, $\tau < 2\tau < 3\tau < \hdots$, of attempt detection before the experiment.
		Here, we choose to measure every $\tau$ units of times.
		In between the measurement times, the system evolves unitarily with the operator $\TEO\sof{\tau} = e^{-i\tau\Ham/\hbar}$, where $\Ham$  is the Hamiltonian of the system.
		The detection attempt is modeled as a strong measurement using the projector $\Detect = \KB{\PsiDet}{\PsiDet}$.
		The detection leads to a collapse of the wave function \cite{Cohen-Tannoudji2009-0}. 
		The detection state $\sKet{\PsiDet}$ is subject to the same restrictions as $\sKet{\PsiIn}$.
		In a lattice system like the tight-binding model, $\sKet{\PsiDet}$ can also be chosen to be a lattice site eigenstate $\sKet{\V{x}_\text{d}}$.

		Directly before the first measurement at time $\tau^-$, the system is in the state 
		\begin{equation}
			\sKet{\psi\sof{\tau^-}} = \TEO\sof{\tau}\sKet{\PsiIn}.
		\label{eq:}
		\end{equation}
		Throughout the manuscript, the superscript $-$($+$) denotes a limit from below(above).
		Now, measurement is attempted, i.e. $\Detect$ is applied.
		The probability to detect the system in $\sKet{\PsiDet}$ in the first attempt is therefore:
		\begin{equation}
			p_1 
			= 
			\sBAK{\psi\sof{\tau^-}}{\Detect}{\psi\sof{\tau^-}}
			=
			\sNorm{\Detect\TEO\sof{\tau}\sKet{\PsiIn}}^2
			,
		\label{eq:}
		\end{equation}
		where $\sNorm{\sKet{\psi}} = \sqrt{\sBK{\psi}{\psi}}$ denotes the usual Hilbert-space norm of a state.
		If the detection was successful, the experiment is finished, and the first detection time is $\tau$.
		In the other case, the wave function collapses (under the orthogonal projection $\Id-\Detect$, $\Id$ being the identity operator), and is renormalized.
		Directly after the first detection attempt, assumed unsuccessful, the wave function is equal to:
		\begin{equation}
			\sKet{\psi\sof{\tau^+}}
			=
			\frac{
				\sbr{\Id-\Detect} \sKet{\psi\sof{\tau^-}}
			}{
				\sqrt{\sBAK{\psi\sof{\tau^-}}{\Id-\Detect}{\psi\sof{\tau^-}}}
			}
			=
			\frac{\sbr{\Id-\Detect} \TEO\sof{\tau} \sKet{\PsiIn}}{\sqrt{1 - p_1}}
			.
		\label{eq:}
		\end{equation}
		For example in a discrete lattice system with $\sKet{\PsiDet} = \sKet{\V{x}_\text{d}}$, the amplitude of the particle at site $\V{x}_\text{d}$ is zero directly after the measurement at time $\tau^+$.
		The paso-doble of unitary evolution and strong measurement is repeated until the first successful detection is registered.
		This ``detection protocol'' combines the collapse of the wave function with the unitary dynamics generated by the Hamiltonian.
		In our setup, detection is attempted stroboscopically, every $\tau$ time units.
		Different choices of the detection times are also possible.
		For example, in Ref.~\cite{Varbanov2008-0} the authors sampled the detection times from a Poisson process.
		Let us assume that the first success occurred at the $n$-th trial.
		Then the wave function directly before this attempt is:
		\begin{equation}
			\sKet{\psi\sof{n\tau^-}}
			=
			\frac{
				\TEO\sof{\tau}\sbrr{\sbr{\Id-\Detect}\TEO\sof{\tau}}^{n-1}\sKet{\PsiIn}
			}{
				\sProd{j=1}{n-1}\sqrt{1-p_j}
			}
			.
		\label{eq:}
		\end{equation}
		The probability to detect the system in this trial {\em under the condition} that it has not been detected before is:
		\begin{equation}
			p_n
			=
			\sBAK{\psi\sof{n\tau^-}}{\Detect}{\psi\sof{n\tau^-}}
			=
			\frac{
				\sNorm{
					\Detect\TEO\sof{\tau}
					\sbrr{\sbr{\Id-\Detect}\TEO\sof{\tau}}^{n-1}
					\sKet{\PsiIn}
				}^2
			}{
				\sProd{j=1}{n-1}\br{1-p_j}
			}
			.
		\label{eq:}
		\end{equation}
		Using the conditional probabilities $p_n$, we can write the unconditioned probability of first detection at the $n$-th attempt as the square norm of some state \cite{Friedman2017-0}:
		\begin{equation}
			F_n
			=
			p_n \Prod{j=1}{n-1} \sbr{1-p_j}
			=
			\sNorm{\Detect\TEO\sof{\tau}\sbrr{\sbr{\Id-\Detect}\TEO\sof{\tau}}^{n-1}\sKet{\PsiIn}}^2
			.
		\label{eq:FnFormulaRefLater}
		\end{equation}
		The non-normalized state on the right hand side is called the detection amplitude \cite{Gruenbaum2013-0,Dhar2015-0}.
		As it is parallel to $\sKet{\PsiDet}$, we can write:
		\begin{equation}
			\varphi_n
			:=
			\sBAK{\PsiDet}{\TEO\sof{\tau}\sbrr{\sbr{\Id-\Detect}\TEO\sof{\tau}}^{n-1}}{\PsiIn}
			.
		\label{eq:DefDetAmp}
		\end{equation}
		The first detection probability is the square norm of this quantity:
		\begin{equation}
			F_n = \sAbs{\varphi_n}^2
		\label{eq:}
		\end{equation}

		It was demonstrated in \cite{Friedman2017-0} that the detection amplitudes defined by Eq.~\eqref{eq:DefDetAmp} obey the following renewal equation:
		\begin{equation}
			\varphi_n
			=
			\sBAK{\PsiDet}{\TEO\sof{n\tau}}{\PsiIn} 
			- \Sum{m=1}{n-1} \sBAK{\PsiDet}{\TEO\sof{\sbr{n-m}\tau}}{\PsiDet} \varphi_m
			.
		\label{eq:QuantumRenewal}
		\end{equation}
		This equation relates the first detection amplitudes with the free evolution of the wave function unperturbed from any measurement.
		The first term is the direct transition from initial to detection state.
		The sum, on the other hand, describes the interference that takes place after the system ``first passed'' the detection state.
		As has been noted in Refs.~\cite{Gruenbaum2013-0,Friedman2017-0}, this equation is formally equivalent to the renewal equation from the first passage theory of random walks, see \cite{Redner2007-0}.
		To obtain the classical equation, one replaces $\varphi_n$ with the first passage probability $F_n^\text{(cl)}$ after $n$ steps.
		Furthermore, $\sBAK{\PsiDet}{\TEO\sof{n\tau}}{\PsiDet}$ is replaced with the probability to return to $\V{x}_\text{d}$ after 		$n$ steps, $P^\text{(cl)}_n\sof{\V{x}_\text{d}|\V{x}_\text{d}}$ and $\sBAK{\PsiDet}{\TEO\sof{n\tau}}{\PsiIn}$ with the probability to move from $\V{x}_\text{in}$ to $\V{x}_\text{d}$ in $n$ steps, $P^\text{(cl)}_n\sof{\V{x}_\text{d}|\V{x}_\text{in}}$ \cite{Redner2007-0}:
		\begin{equation}
			F_n^\text{(cl)} 
			=
			P_n^\text{(cl)}\sof{\V{x}_\text{d}|\V{x}_\text{in}}
			-
			\Sum{m=1}{n-1} 
			P_{n-m}^\text{(cl)}\sof{\V{x}_\text{d}|\V{x}_\text{d}}
			F_m^\text{(cl)} 
			.
		\label{eq:ClassicalRenewal}
		\end{equation}

		Just like in random walk theory, we solve the equation with generating functions:
		\begin{align}
			\label{eq:DefZDetAmp}
			\varphi\of{z}	:= &	\Sum{n=1}{\infty} z^n \varphi_n \\
			\Resolv\of{z} 		:= &	\sSum{n=0}{\infty} z^n \TEO\sof{n\tau}
			.
			\label{eq:DefResolv}
		\end{align}
		Observe that we put $\varphi_0 := 0$, since the first detection attempt happens at $1\times\tau$.
		The generating function of the evolution operator is closely related to its resolvent $\Resolv\sof{z} = z^{-1} \sbrr{ z^{-1} - \TEO\sof{\tau}}^{-1}$.
		Eq.~\eqref{eq:QuantumRenewal} is now multiplied with $z^n$ and summed from $n=1$ to infinity.
		One obtains \cite{Friedman2017-0,Friedman2017-1}:
		\begin{equation}
			\varphi\of{z}
			=
			\frac{ 
				\BAK{\PsiDet}{\Resolv\of{z}}{\PsiIn}
				- \BK{\PsiDet}{\PsiIn} 
			}{
				\BAK{\PsiDet}{\Resolv\of{z}}{\PsiDet}
			}
			.
		\label{eq:QuantumRenewalSol2}
		\end{equation}

		For notational simplicity, we will henceforth consider the {\em return} problem only.
		That means we only consider $\sKet{\PsiDet} = \sKet{\PsiIn}$ in the main text.
		We stress that the derivation of the arrival problem (i.e. $\sKet{\PsiDet} \ne \sKet{\PsiIn}$) follows exactly along the same lines.
		This is demonstrated in Appendix~\ref{app:Arrival}, where we explain the necessary modifications.
		Asymptotically, Eq.~\eqref{eq:AsymFDP} is valid for the return as well as the arrival problem, although both have different amplitudes, $F_{l,d_S}$.
		Consequently, any relation between asymptotic first detection probabilities and the distance between starting and final position appears in these amplitudes, and not in the exponents or frequencies \cite{Thiel2018-0}.

		Let us fix our language and notation.
		We abbreviate:
		\begin{align}
			\label{eq:DefRA}
			u_n := & \sBAK{\PsiDet}{\TEO\sof{n\tau}}{\PsiDet}, \\
			u\sof{z} := & \Sum{n=0}{\infty} u_n z^n = \sBAK{\PsiDet}{\Resolv\sof{z}}{\PsiDet}
			.
			\label{eq:DefResolv}
		\end{align}
		We refer to $u_n$ as the return amplitude and to $u\sof{z}$ as the resolvent.
		``The'' generating function refers to $\varphi\sof{z}$.
		With these symbols and with the normalization $\sBK{\PsiDet}{\PsiDet} = 1$, we can rewrite Eq.~\eqref{eq:QuantumRenewalSol2} as \cite{Friedman2017-1}:
		\begin{equation}
			\varphi\sof{z}
			=
			1 - \frac{1}{u\sof{z}}
			.
		\label{eq:GenFunc}
		\end{equation}

		The $z$-transform can be inverted using Cauchy's integral formula:
		\begin{equation}
			\varphi_n
			=
			\frac{1}{2\pi i} \ointctrclockwise\limits_{\Abs{z}=r} \frac{\D z}{z^{n+1}}
			\brr{ 
				1 
				-
				\frac{1}{
					u\sof{z}
				}
			}
			.
		\label{eq:CauchyInt}
		\end{equation}
		$r\le1$ is the radius of the circle contour that only contains the pole at the origin.
		(There is no other pole inside the unit circle, because $\varphi\sof{z}$ is by definition analytic inside the unit disk.)
		Now, the integration contour is parametrized as $z=re^{i\lambda}$ and the limit $r\to1^-$ is taken:
		\begin{equation}
			\varphi_n
			=
			\Int{0}{2\pi}{\lambda} \frac{e^{-in\lambda}}{2\pi}
			\brr{ 1 - \frac{1}{u\sof{e^{i\lambda}}} }
			.
		\label{eq:FourierInt}
		\end{equation}
		This identifies $\varphi_n$ as a Fourier transform, for which asymptotic formulae are readily available \cite{Gamkrelidze1989-0,Erdelyi1956-0,Cline1991-0}.
		The large $n$ asymptotics of $\varphi_n$ are related to the singularities $\Lambda^*_l$, $l\in\sbrrr{0,\hdots,L-1}$, at which $\varphi\sof{e^{i\lambda}}$ is non-analytic as a function of $\lambda$.
		The resolvent of an operator is a standard tool to infer an operator's spectrum.
		Since $u\sof{z}$ is equal to the resolvent of the evolution operator up to some factor, its properties are determined by the {\em energy spectrum}.
		Consequently, the detection amplitude's properties are determined by the energy spectrum as well.
		We will restrict ourselves to systems with a continuous energy spectrum.
		This allows us to express the integrand $\varphi\sof{e^{i\lambda}}$ in terms of the so-called MSDOS, which is itself related to the DOS.
		This is the subject of the next section.

	\section{The density of energy states and the measurement spectral density of states}
	\label{sec:SpecMeas}
		\subsection{The density of states and the spectral dimension}
			Before treating the continuous spectrum case, let us first consider a finite system of Hilbert space dimension $N$ with time independent Hamiltonian $\Ham$.
			The (possibly degenerate) eigen-energies are $E_n$ and the corresponding eigenstates are $\sKet{\chi_{n,j}}$, where $j$ enumerates the degeneracy.
			The DOS is defined by:
			\begin{equation}
				\rho\sof{E}
				:= 
				\frac{1}{N}
				\Sum{n,j}{} \delta\sof{E-E_n}
				=
				\frac{1}{N} \Trace{\delta\sof{E - \Ham}}
				.
			\label{eq:DOSFinite}
			\end{equation}
			In a system with finite-dimensional Hilbert space, $\rho\sof{E}$ is always a sum of delta functions.
			When the thermodynamic limit $N\to\infty$ is taken, $\rho\sof{E}$ can become a mixture of delta functions and a density function.
			(Instead of $N$, any extensive quantity can be used to normalize the thermodynamic limit.)
			In this manuscript, we deal with systems that have no discrete energy states so that $\rho\sof{E}$ contains no delta functions.

			An example is the tight-binding Hamiltonian of Eq.~\eqref{eq:TBHamiltonian}, which is diagonalized by free wave states $\sKet{\V{k}} := \sbr{a/(2\pi)}^{d/2} \sSum{\V{x}\in a \Integers}{} e^{i \V{k}\V{x} }$, where $\V{k}$ is taken from the cube-shaped Brillouin zone $\Brillouin=[-\pi/a,\pi/a]^d$.
			The dispersion relation is the relation between energy and wave-vector:
			\begin{equation}
				E\of{\V{k}}
				:=
				4 \gamma 
				\Sum{j=1}{d} 
				\sin^2\sof{\tfrac{ak_j}{2}}
				,
				\label{eq:TBDispersion}
			\end{equation}
			where $k_j$ is the $j$-th component of the vector $\V{k}$.
			The corresponding DOS can be obtained by an integral over the Brillouin zone.
			In one dimension, the DOS is given by the arcsin law:
			\begin{equation}
				\rho\sof{E} 
				=
				\frac{a}{2\pi}
				\Int{-\frac{\pi}{a}}{\frac{\pi}{a}}{k}\delta\sof{E - E\sof{k}}
				=
				\frac{1}{\pi}
				\frac{1}{\sqrt{\sbr{4\gamma - E} E}}
				.
			\label{eq:TBDOS}
			\end{equation}
			This is seen from plugging Eq.~\eqref{eq:TBDispersion} into the middle expression of Eq.~\eqref{eq:TBDOS} and changing the integration variable.
			The allowed energies lie in the interval $[0,4\gamma]$; outside this interval, we have $\rho\sof{E} = 0$.
			The DOS in higher dimensions can be interpreted as the PDF of the sum of random variables $E\sof{\V{k}}$ given by Eq.~\eqref{eq:TBDispersion}, when the $k_j$'s are thought of as i.i.d. random variables that are uniformly distributed in $[-\pi/a,\pi/a]$.
			Consequently, $\rho\sof{E}$ for higher dimensions can be represented as a convolution integral:
			\begin{equation}
				\rho^{(d)}\sof{E}
				=
				\Int{}{}{E'}
				\frac{1}{\pi}
				\frac{1}{\sqrt{\sbr{4\gamma - E'} E'}}
				\rho^{(d-1)}\sof{E-E'}
				,
			\label{eq:}
			\end{equation}
			where $\rho^{(1)}\sof{E}$ is given by Eq.~\eqref{eq:TBDOS}.
			In two dimensions $\rho\sof{E}$ can be expressed in terms of complete elliptic integrals \cite{Montroll1947-0}.

			It is apparent from Eq.~\eqref{eq:TBDOS} that the DOS is not analytic everywhere.
			At the singular energies $E^*_0 = 0$ and $E^*_1 = 4\gamma$, it exhibits a transition from one-over-square-root to vanishing behavior.
			These singular points are called the van Hove singularities \cite{vanHove1953-0} and are a consequence of differential geometric considerations \cite{Arnold2012-1}.
			One of these points is always located at the lowest possible energy (which we fix at zero).
			The total number of non-analytic points in $\rho\sof{E}$ depends on the system at hand.
			In higher dimensions there may be more than two of those points and the singularity may be present in some derivative of $\rho\sof{E}$.
			This behavior is generic and defines the so-called {\em spectral dimension}, $d_S^\text{DOS}$, of the system:
			The non-analytic term in $\rho\sof{E}$ behaves as $\sAbs{E-E^*}^{d_S^\text{DOS}/2-1}$ around the singular point $E^*$ \cite{Alexander1982-0}.
			Logarithmic factors appear in even dimensions \cite{vanHove1953-0,Maradudin1958-0,Arnold2012-0}.
			In certain systems with fractal characteristics, the spectral dimension can differ from the Euclidean one \cite{Hughes1995-0}.

\begin{figure*}
				\includegraphics[width=0.99\textwidth]{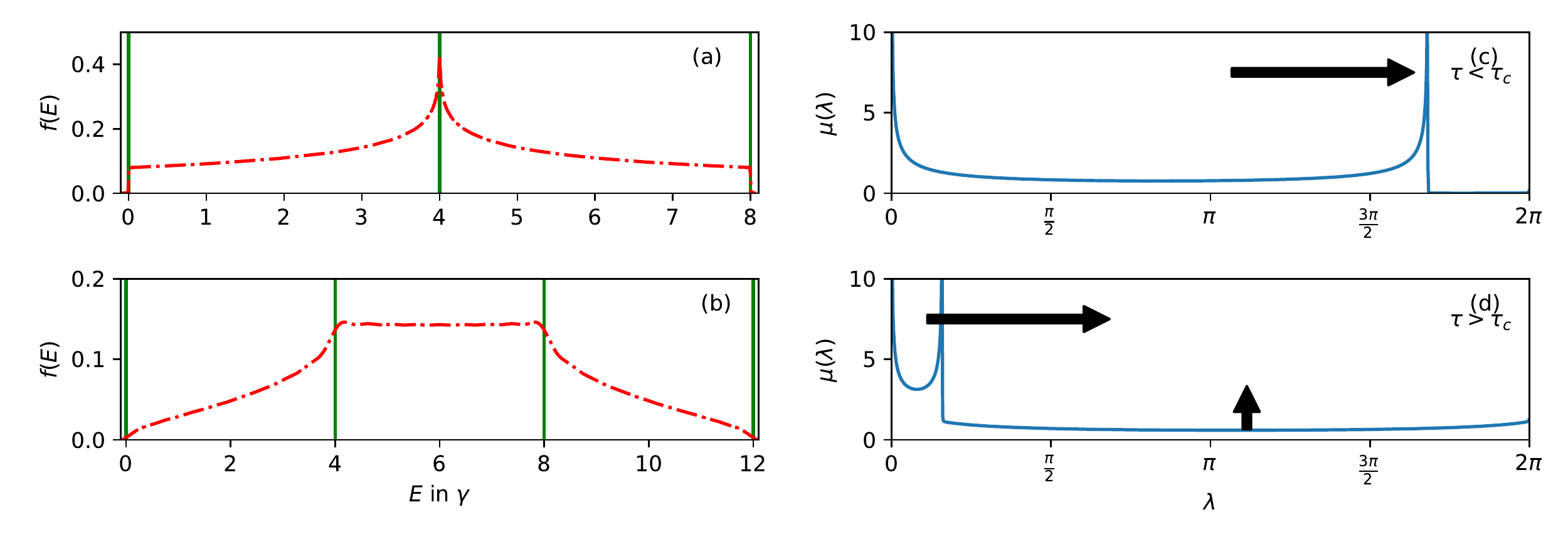}
				\caption{
					MSDOS and WMSDOS for the tight-binding model with detection at the origin.
					Left: The MSDOS $f\sof{E}$ of the tight-binding model.
					For our special choice of detection and initial states, $f\sof{E}$ is equal to the DOS $\rho\sof{E}$.
					(a): $f\sof{E}$ for $d=2$.
					There are three singular energies at $0$, $4\gamma$, and $8\gamma$, indicated by the vertical lines.
					At the outer singularities, $f\sof{E}$ is discontinuous; it vanishes outside $[0,8\gamma]$.
					In the middle there is a logarithmic divergence.
					(b): $f\sof{E}$ for $d=3$.
					There are four singularities in the derivative of $f\sof{E}$ at multiples of $4\gamma$ indicated by the vertical lines.
					Around the singularities $f\sof{E}$ behaves like Eq.~\eqref{eq:SpecMeasAssump}.
					Compare these figures with the DOS sketches of \cite{Maradudin1958-0}.
					Right: $\mu\sof{\lambda}$ for the one dimensional tight-binding model, Eq.~\eqref{eq:TBWMSDOS}.
					(c): For the detection period smaller than the critical value $\tau < \tau_c = (\pi/2) \hbar/\gamma$.
					The two singularities are clearly visible.
					The right singularity moves further to the right as $\tau$ increases and merges with the left one at the critical value.
					(Note that $\mu\sof{\lambda}$ is $2\pi$ periodic.)
					Also $\mu\sof{\lambda}$ vanishes outside the singularities because the whole support of $f\sof{E}$ is mapped into the interval $[0,2\pi]$.
					(d): $\mu\sof{\lambda}$ for $\tau>\tau_c$.
					The second singularity reappears on the left hand side of the plot for $\tau >\tau_c$.
					It moves to the right for growing $\tau$ until $\tau$ is equal to another multiple of the critical value.
					Notice that $\mu\sof{\lambda}$ does not vanish outside the singularities, because $f\sof{E}$ is wrapped more than once around the unit circle.
					The non-vanishing area corresponds to additional terms in the sum of Eq.~\eqref{eq:TBWMSDOS} that appear when $\tau$ is larger than $\tau_c$.
					\label{fig:SM}
				}
			\end{figure*}

			The DOS for the two and three dimensional tight-binding model is plotted in Fig.~\ref{fig:SM}(a-b).
			Our plots can be compared with the sketches from \cite{Maradudin1958-0}.
			We marked the position of the van Hove singularities by vertical lines in Fig.~\ref{fig:SM}.

			For all translational invariant systems, the DOS can be obtained from an integral over the Brillouin zone using the system's dispersion relation.
			The latter defines a surface in $\sbr{\V{k},E}$-space.
			The van Hove singularities are related to the critical points of the surface \cite{Arnold2012-1}.
			Their maximal number is a topological property of this surface and is related to its Betti number \cite{vanHove1953-0}.
			For a lattice system, the dispersion relation is always periodic over a Brillouin-zone; hence the energy surface is a torus that has multiple critical points.
			For a particle in continuous space, the dispersion relation is a parabola, which has only one critical point.

		\subsection{The measurement spectral density of states}
			The DOS is a property of the Hamiltonian only, and encodes no information about the detection state.
			A more precise tool is needed in our situation: the measurement spectral density of states (MSDOS) associated with the state $\sKet{\PsiDet}$.
			Instead of taking the trace of the operator $\delta\sof{E-\Ham}$, as we did in Eq.~\eqref{eq:DOSFinite}, we will now only consider one matrix element.
			\begin{equation}
				f\sof{E}
				:=
				\BAK{\PsiDet}{\delta\sof{E-\Ham}}{\PsiDet}
				=
				\Sum{n,j}{} \sAbs{\sBK{\chi_{n,j}}{\PsiDet}}^2
				\delta\sof{E-E_n}
				.
			\label{eq:SpecMeasFinite}
			\end{equation}
			$f\sof{E}$ can be thought of as the squared modulus of $\sKet{\PsiDet}$'s ``energy representation''.
			Just as before, $f\of{E}$ will approach a function of $E$ without any delta-function contributions in the thermodynamic limit.
			It is normalized to unity, $\sInt{0}{\infty}{E} f\sof{E} = \sBK{\PsiDet}{\PsiDet} = 1$. 
			(Remember that the lowest possible energy is fixed at zero.)
			In the mathematical literature \cite{Marchetti2012-0}, it is known as the Hamiltonian's spectral measure associated with the state $\sKet{\PsiDet}$.
			If $\sKet{\PsiDet}$ is an atomic orbital state or some other spatially localized wave function, $f\sof{E}$ can also be identified with the local DOS \cite{Busch1998-0,Li2001-0,Yeganegi2014-0}.

			The advantage of the MSDOS is that matrix elements of the Hamiltonian or related operators can be represented as integrals over all energies, in particular the return amplitudes:
			\begin{equation}
				u_n 
				= 
				\BAK{\PsiDet}{e^{-in\frac{\tau\Ham}{\hbar}}}{\PsiDet}
				=
				\Int{0}{\infty}{E} e^{-in\frac{\tau E}{\hbar}} f\sof{E}
				.
			\label{eq:RA}
			\end{equation}
			The return amplitude and the MSDOS are a Fourier transform pair.

			Consider the case, when the system at hand is invariant under a certain symmetry transformation.
			Then one may be able to find a set of base states $\sKet{\tilde{\chi}_n}$, such that $\sBAK{\tilde{\chi}_n}{\Ham}{\tilde{\chi}_n} = \sBAK{\tilde{\chi}_{n'}}{\Ham}{\tilde{\chi}_{n'}}$, for any pair $n$,$n'$.
			In this case $N^{-1} \sTrace{\delta\sof{E - \Ham}} = \sBAK{\tilde{\chi}_n}{\delta\sof{E-\Ham}}{\tilde{\chi}_n}$ for any $n$.
			This means that the MSDOS associated with $\sKet{\tilde{\chi}_n}$ {\em can be identified with the DOS.}

			This is exactly the situation in the tight-binding model, for the special case when $\sKet{\PsiDet}$ is a lattice site eigenstate $\sKet{\V{x}_\text{d}}$:
			By translational invariance, the matrix elements $\sBAK{\V{x}}{\Ham}{\V{y}}$ only depend on the distance $\V{y}-\V{x}$.
			In one dimension, one obtains with Eq.~\eqref{eq:TBDOS}:
			\begin{equation}
				f\sof{E} = \rho\sof{E}
				=
				\frac{1}{\pi}
				\frac{1}{\sqrt{E\sof{4\gamma-E}}}
			\label{eq:TBSM}
			\end{equation}
			Consequently, the van Hove singularities and the spectral dimension can be found in $f\sof{E}$ as well.

			In general, $f\sof{E}$ and $\rho\sof{E}$ are not equal!
			As a counterexample, consider the detection state $\sKet{\PsiDet} = \sbr{\sKet{a} + \sKet{-a}}/\sqrt{2}$ and compute its MSDOS from the Brillouin zone integral with $\sBK{\pm a}{k} = \sqrt{a/(2\pi)} e^{\pm i a k}$:
			\begin{align}
				f\sof{E}
				= & \nonumber 
				\frac{1}{2}
				\Int{-\frac{\pi}{a}}{\frac{\pi}{a}}{k}
				\delta\sof{E - E\sof{k}} 
				\br{\Bra{a} + \Bra{-a}}
				\KB{k}{k}
				\br{\Ket{a} + \Ket{-a}}
				\\ = &
				\frac{a}{4\pi} \Int{-\frac{\pi}{a}}{\frac{\pi}{a}}{k}
				\delta\sof{E - E\sof{k}}
				4 \cos^2\sof{ak}
				=
				\frac{1}{\pi} \frac{\br{1-\frac{E}{2\gamma}}^2}{\sqrt{E\sbr{4\gamma-E}}}
				.
				\label{eq:TBSMCounterEx1}
			\end{align}
			Thus, here the MSDOS has an additional factor relative to the DOS.
			Although the MSDOS and the DOS are different for this detection state, we note that they both feature a one-over-square-root divergence at $E=0$ and at $E=4\gamma$.
			That is, the non-analytic points of $f\sof{E}$ are the van Hove singularities and the spectral dimension $d_S$ is the same as the one found in the DOS $d_S^\text{DOS}$.

			As we have shown, $f\sof{E}$ and $\rho\sof{E}$ are not equal for a general choice of detection state.
			Nevertheless, $f\sof{E}$ may have $L'$ non-analytic points $E^*_l$, just like $\rho\sof{E}$ has the van Hove singularities.
			We assume that $f\sof{E}$ admits the following asymptotic expansion around these points:
			\begin{equation}
				f\sof{E^*_l\pm\epsilon}
				\AsymEq
				\tilde{f}_l\sof{\pm\epsilon}
				+
				A^\pm_l \epsilon^{\frac{d_S}{2}-1}
				,
			\label{eq:SpecMeasAssump}
			\end{equation}
			where the coefficients $A^\pm_l$ depend on the particular point $E^*_l$ and on the direction of the approach.
			Since the singularity may be present only in a derivative, we introduce the ``analytic remainder'', which is nothing else but the Taylor expansion up to order $d_S/2-1$:
			\begin{equation}
				\tilde{f}_l\sof{\pm\epsilon}
				=
				\Sum{0\le n < \tfrac{d_S}{2}-1}{} 
				\frac{f^{(n)}\sof{E^*_l}}{n!} \sbr{\pm\epsilon}^n
				.
			\label{eq:DefAnalRemain}
			\end{equation}
			Clearly, an expansion like Eq.~\eqref{eq:SpecMeasAssump} is always possible.
			However, the identification of $f\sof{E}$'s singularities with those of $\rho\sof{E}$ is not always possible.
			We consider a detection state as ``ordinary'' when the corresponding MSDOS's singularities are located where the van Hove singularities are, and when its spectral dimension coincides with $d_S^\text{DOS}$.

			For the 1d tight-binding model with $\sKet{\PsiDet} = \sKet{x_\text{d}}$, we see from  Eq.~\eqref{eq:TBSM} that 
			\begin{align}
				f\sof{0+\epsilon} = & f\sof{4\gamma - \epsilon} = \frac{1}{\pi\sqrt{4\gamma}} \epsilon^{\frac{1}{2}-1}	\\
				f\sof{0-\epsilon} = & f\sof{4\gamma + \epsilon} = 0.
			\end{align}
			This identifies the spectral dimension as unity and $A_0^- = A_1^+ = 0$ as well as $A_0^+ = A_1^- = (4 \pi^2 \gamma)^{-1/2}$ and $\tilde{f}_l\sof{\pm\epsilon} = 0$.
			Table~\ref{tab:Constants} lists all the constants used throughout the manuscript.
			As mentioned, when the detection state is chosen as a lattice site eigenstate, $f\sof{E}$ and $\rho\sof{E}$ will coincide in the tight-binding model in any dimension.
			\begin{table}
				\begin{tabular}{c||c|c|c|c}
								&	TB	&	FP	&	L\'evy	&	Def.					\\
								\hline \hline
					$L$			&	$d+1$	&	$1$	&	$1$	&	Sec.~\ref{sec:SpecMeas}	\\
					$E^*_l$		&	$4 l \gamma$	&	$0$		&	$0$	&	Eq.~\eqref{eq:SpecMeasAssump}						\\
					$d_S$		&	$d$	&	$d$	&	$2\frac{d}{\alpha}$		&	Eq.~\eqref{eq:SpecMeasAssump}	\\  
					$A^+_l$		&	$\frac{\sbr{-1}^{\frac{l}{2}}\binom{d}{l}}{\sGma{\frac{d}{2}}\sbr{4\pi\gamma}^\frac{d}{2}}$	&	$\frac{E_0^{-\frac{d}{2}}}{\Gma{\frac{d}{2}}} $	&	$\frac{2}{\alpha}\frac{E^{-\frac{d}{\alpha}}_0}{\Gma{\frac{d}{2}}} $	&	Eq.~\eqref{eq:SpecMeasAssump}					\\
					$A^-_l$		&	$\frac{\sbr{-1}^{\frac{d-l}{2}}\binom{d}{l}}{\sGma{\frac{d}{2}}\sbr{4\pi\gamma}^\frac{d}{2}}$	&	$0$	&	$0$		&	Eq.~\eqref{eq:SpecMeasAssump}				\\
					$C_l$	&	$i^l \binom{d}{l} \br{\frac{-i\hbar}{4\pi\gamma\tau}}^{\frac{d}{2}}$	&	$\br{\frac{-i\hbar}{E_0\tau}}^{\frac{d}{2}}$	&	$\frac{\Gma{1+\frac{d}{\alpha}}}{\Gma{1+\frac{d}{2}}} \br{\frac{-i\hbar}{E_0\tau}}^{\frac{d}{\alpha}}$	&	Eq.~\eqref{eq:DefC}	\\
					\hline
					Eq.			&	\eqref{eq:DefCTB}	&	\eqref{eq:NRA}	&	\eqref{eq:LevyA} & 
				\end{tabular}
				\caption{
					Table of the different coefficients in the tight-binding model (TB), the free particle (FP) and the L\'evy particle (L\'evy).
					In the tight-binding model the detection state is a lattice site eigenstate $\sKet{\PsiDet} = \sKet{\V{x}_\text{d}}$.
					For the other two models the detection state is a Heisenberg state given by Eq.~\eqref{eq:NRDefPsi}.
					$0 < \alpha < 2$ is the L\'evy parameter, see Eq.~\eqref{eq:LevyDisp}.
					The last column lists where to find the definition of the quantity.
					The last row lists, where the specific result is found in the main text.
					In the continuous space models, we used Eq.~\eqref{eq:DefC} to compute the constants $C_l$ from the $A^\pm$'s.
					In the tight-binding model, we also used Eq.~\eqref{eq:DefC} to obtain the $A^\pm$'s.\footnote{
						Eq.~\eqref{eq:DefC} alone is not sufficient to determine both $A^+_l$ and $A^-_l$ from $C_l$, because it is one equation for two variables.
						We employed the additional condition, that $A^\pm_l$ must be real from which we inferred that $A^\pm_l$ vanishes for some $l$.
						A rigorous computation involves the Mellin transform of the MSDOS around one of its singular points and uses its representation as an integral over the Brillouin zone \cite{Arnold2012-1}:
						\begin{equation*}
							\mathcal{M}_l^\pm\sofff{f;s} 
							:= 
							\Int{0}{\infty}{\epsilon} 
							\epsilon^{s-1}
							\Int{\Brillouin}{}{\V{k}} 
							\delta\sof{E^*_l \pm \epsilon - E\sof{\V{k}}}
							\sAbs{\PsiDet\sof{\V{k}}}^2
							,
						\end{equation*}
						where $\PsiDet\sof{\V{k}}$ is the momentum representation of $\sKet{\PsiDet}$.
						The delta function is easily resolved, and the remaining integrand is expanded up to second order around the critical points $\V{k}^*$ {\em of the energy surface} $E\sof{\V{k}}$ that correspond to the singular energy $E^*_l$, i.e. $E\sof{\V{k}^*} = E^*_l$.
						The Mellin transform has several poles in the complex $s$-plane. 
						The pole with the largest real part lies at $s=d_S/2-1$ and determines the small $\epsilon$ behavior of the MSDOS.
						The coefficients $A^\pm_l$, as well as possible logarithmic factors can be extracted from this pole.
						This is possible for any translationally invariant system.
						A full derivation will be carried out in another publication.
					}
					The tight-binding entry for $A^+_l$ is valid for even $l$ and zero otherwise.
					The tight-binding entry for $A^-_l$ is valid for even $d-l$ and zero otherwise.
					In even dimensions $A^-_0$ and $A^+_d$ vanish and furthermore logarithmic factors appear.
					For the free particle and the L\'evy particle there is only one singular point with $l=0$.
					\label{tab:Constants}
				}
			\end{table}
			In this special case the return amplitudes are Bessel functions of the first kind.
			This can again be seen from an integral over the Brillouin zone using the dispersion relation Eq.~\eqref{eq:TBDispersion}:
			\begin{equation}
				u_n 
				= 
				\frac{a}{2\pi}
				\Int{-\frac{\pi}{a}}{\frac{\pi}{a}}{k}
				e^{-i 2n\tfrac{\gamma\tau}{\hbar}\br{1 - \cos\sof{ak}}}
				=
				e^{-i2n\tfrac{\gamma\tau}{\hbar}}
				\BesselJ{0}{2n\tfrac{\gamma\tau}{\hbar}}
				,
			\label{eq:RATB1D}
			\end{equation}
			where we first used the replacement $2\sin^2\sof{x/2} = 1 - \cos x$, and then the integral representation of the Bessel function.
			In higher dimensions of the tight-binding model, the integral over the Brillouin zone factorizes and we find:
			\begin{equation}
				u_n 
				= 
				\Prod{j=1}{d}
				\frac{a}{2\pi}
				\Int{-\frac{\pi}{a}}{\frac{\pi}{a}}{k_j}
				e^{-i n\tfrac{4\gamma\tau}{\hbar}\sin^2\tfrac{ak_j}{2}}
				=
				\brr{
				e^{-in\tfrac{2\gamma\tau}{\hbar}}
				\BesselJ{0}{\tfrac{2\gamma\tau}{\hbar}n}}^d
			\label{eq:RATB}
			\end{equation}
			An alternative integral representation of the Bessel function [Eq.~8.411(10) of \cite{Gradshteyn2007-0}] reveals the Bessel function as the Fourier transform of the arcsin law.
			This allows us to use Eq.~\eqref{eq:TBSM} directly in Eq.~\eqref{eq:RA} in the 1d case:
			\begin{equation}
				u_n 
				= 
				\frac{1}{\pi} 
				\Int{0}{4\gamma}{E} 
				\frac{e^{-in\frac{\tau E}{\hbar}}}{\sqrt{\sbr{4\gamma - E}E}}
				=
				e^{-i\tfrac{2\gamma\tau}{\hbar}n}
				\BesselJ{0}{\tfrac{2\gamma\tau}{\hbar}n}
				,
			\label{eq:}
			\end{equation}
			where the variable change $E = 2\gamma(1+x)$ has to be used to recover the reference's formula.
			We plot the MSDOS of the tight-binding model for two and three dimensions in Fig.~\ref{fig:SM}(a-b).

		\subsection{The wrapped MSDOS}
			In Eq.~\eqref{eq:RA}, we expressed the return amplitudes in terms of $f\sof{E}$.
			The same can be done to the resolvent [using Eq.~\eqref{eq:RA}, Eq.~\eqref{eq:DefResolv} and the geometric series]:
			\begin{equation}
				u\sof{z}
				=
				\Int{0}{\infty}{E}
				\frac{f\sof{E}}{1 - z e^{-i \frac{\tau E}{\hbar}}}
				=
				\frac{1}{2\pi} \Int{0}{2\pi}{\lambda'}
				\frac{\mu\sof{\lambda'}}{1 - ze^{-i\lambda'}}
				.
			\label{eq:DefResolvWMSDOS}
			\end{equation}
			Since the complex exponential in the denominator is periodic, it makes sense to gather all contributions of $f\sof{E}$ with the same phase.
			The result is the ``wrapped MSDOS'' (WMSDOS):
			\begin{equation}
				\mu\of{\lambda} 
				:= 
				\frac{2\pi \hbar}{\tau}
				\Sum{m=-\infty}{\infty} 
				f\of{\tfrac{\hbar}{\tau}\sbrr{\lambda+2\pi m}}
				.
			\label{eq:DefWMSDOS}
			\end{equation}
			$\mu\sof{\lambda}$ can be understood as ``$f\sof{E}$ wrapped around the unit circle''.
			It is actually the spectral measure of the evolution operator associated with $\sKet{\PsiDet}$.
			It is normalized according to $(2\pi)^{-1}\sInt{0}{2\pi}{\lambda}\mu\sof{\lambda} = \sBK{\PsiDet}{\PsiDet} = 1$.
			Example plots of $\mu\sof{\lambda}$ can be found in the insets of Fig.~\ref{fig:SM}.
			In the mathematical literature, Eq.~\eqref{eq:DefResolvWMSDOS} is called the Cauchy transform of the measure $\mu\of{\lambda}\D \lambda$ \cite{Cima2006-0}.
			In contrast to the series definition, Eq.~\eqref{eq:DefResolv}, the integral representation is also valid for $\sAbs{z} > 1$.

			A system with an infinite energy band, like a free particle in continuous space, will always have infinitely many terms in the sum of Eq.~\eqref{eq:DefWMSDOS}.
			For a system with a finite energy band, like the tight-binding model, most of the terms in Eq.~\eqref{eq:DefWMSDOS} will be zero, because they are outside the support of $f\sof{E}$.
			The support of $f\sof{E}$ gets stretched or compressed by a factor $\tau/\hbar$ before it is wrapped onto the interval $[0,2\pi]$.

			At certain critical values of $\tau$ a new term will appear in the sum of Eq.~\eqref{eq:DefWMSDOS}.
			To better understand Eq.~\eqref{eq:DefWMSDOS}, consider the one dimensional tight-binding model again.
			For very small values of $\tau$, smaller than the critical value:
			\begin{equation}
				\tau_c
				:=
				\frac{2\pi\hbar}{4\gamma} 
				= 
				\frac{\pi\hbar}{2\gamma}
				,
			\label{eq:TBDefCritTau}
			\end{equation}
			which is set by the width of the energy band, $4\gamma$ [see Eq.~\eqref{eq:TBDispersion}], the support of $\mu\sof{\lambda}$ is actually smaller than $2\pi$ and $\mu\sof{\lambda}$ is just a rescaled version of $f\sof{E}$.
			Additional terms appear in $\mu\sof{\lambda}$, as soon as $\tau$ surpasses a multiple of the critical value $\tau_c$.
			Assume that $\sbr{n-1}\tau_c < \tau < n\tau_c$ for some positive integer $n$, then we obtain by combining Eq.~\eqref{eq:DefWMSDOS} and Eq.~\eqref{eq:TBSM}:
			\begin{equation}
				\mu\sof{\lambda}
				=
				\Sum{m=0}{n}
				\frac{2}{\sqrt{
					\sbr{\lambda+2\pi m} 
					\br{\tfrac{4 \gamma\tau}{\hbar} - \lambda - 2\pi m}
				}}
				.
			\label{eq:TBWMSDOS}
			\end{equation}
			We see that $f\sof{E}$'s singularities are inherited by $\mu\sof{\lambda}$.

			The critical behavior of $\mu\sof{\lambda}$ around its singular points translates to the behavior of the resolvent $u\sof{re^{i\lambda}}$ close to the unit circle, i.e. in the limit $r\to1^-$.
			As we show in Appendix~\ref{app:Plemelj}, the chain of definitions for $f\sof{E}$ from Eq.~\eqref{eq:SpecMeasAssump} can be traced forward to $u\sof{e^{i\lambda}}$, in order to find the singularities in the resolvent.
			We summarize this behavior in the following equation:
			\begin{equation}
				u\sof{e^{i\sbr{\Lambda^*_l \pm\epsilon}}}
				\AsymEq
				\tilde{u}_l\sof{\pm\epsilon}
				+
				B^\pm_l\epsilon^{\frac{d_S}{2}-1}
				,
			\label{eq:AsymResolv}
			\end{equation}
			where the notation is similar to Eq.~\eqref{eq:SpecMeasAssump}.
			The constants $B^\pm_l$ and $A^\pm_l$ are related, as we will show later in Eqs.~\eqref{eq:DefC} and \eqref{eq:DefCTB}, where both are computed from information about $u_n$.
			The particular way how one obtains these constants -- from $f\sof{E}$, or from $u_n$ -- is a matter of convenience.

			The wrapping procedure Eq.~\eqref{eq:DefWMSDOS} shifts the positions of the singularities from $E^*_l$ to:
			\begin{equation}
				\Lambda^*_l 
				:= 
				\frac{E^*_l \tau}{\hbar}
				\mod 2\pi
				.
			\label{eq:DefSingLambdas}
			\end{equation}
			For the 1d tight-binding model with localized detection state, these are the points
			\begin{equation}
				\Lambda^*_0 = 0, \qquad \Lambda^*_1 = \frac{4\gamma\tau}{\hbar} \mod 2\pi
				.
			\label{eq:}
			\end{equation}
			For special choices of $\tau$, two singular energies $E^*_l$ and $E^*_{l'}$ become equivalent:
			\begin{equation}
				\tau_c^{(l,l')}
				:=
				\frac{2 \pi \hbar}{\sAbs{E^*_l - E^*_{l'}}}
				.
			\label{eq:DefCritTau}
			\end{equation}
			These are the critical sampling periods \cite{Thiel2018-0}, and we will later show that such choices of $\tau$ yield special behavior of the first detection probabilities.
			In the tight-binding model these are:
			\begin{equation}
				\tau_c^{(l)}
				=
				\frac{\pi\hbar}{2l\gamma}
				,
			\label{eq:TBDefCritTaus}
			\end{equation}
			for $l \in \sbrrr{1,\hdots,d}$.
			At these critical detection periods, two or more singularities of $f\sof{E}$ get mapped to {\em one} singularity of $\mu\sof{\lambda}$.
			The number $L$ of singularities of $\mu\sof{\lambda}$ is then smaller than $L'$, which is the number of singularities in $f\sof{E}$.
			In the one dimensional tight-binding model, this is the already encountered critical value from Eq.~\eqref{eq:TBDefCritTau}.
			However, $\mu\sof{\lambda}$'s singularities exhibit exactly the same power laws as $f\sof{E}$.

		\subsection{Singularities in the generating function}
			We conclude this section with identifying the singularities in the generating function of the detection amplitudes evaluated on the unit circle.
			This is done by taking the limit $r\to1^-$ in Eq.~\eqref{eq:GenFunc} and using Eq.~\eqref{eq:AsymResolv}.
			The singularities of $\varphi\sof{e^{i\lambda}}$ are the points $\Lambda^*_l$ defined by Eq.~\eqref{eq:DefSingLambdas}.
			Close to these points, we have:
			\begin{equation}
				\varphi\sof{e^{i\sbr{\Lambda^*_l+\epsilon}}}
				\AsymEq
				1 -
				\frac{1}{
					\tilde{u}_l\sof{\pm\epsilon}
					+
					B^\pm_l\epsilon^{\frac{d_S}{2}-1}
				}
				.
			\end{equation}
			We find a competition of terms in the denominator.
			Depending on the value of the spectral dimension, either the power term or the analytic remainder dominates.
			For $d_S\le2$ the analytic remainder $\tilde{u}_l$ is zero, while for $d_S >2$ it constitutes the leading order.
			Performing the small-$\epsilon$ expansion, we find that $\varphi\sof{e^{i\lambda}}$'s singularity is always in one of its derivatives.
			There is a crossover at the critical dimension $d_S = 2$:
			\begin{equation}
				\varphi\sof{e^{i\sbr{\Lambda^*_l\pm\epsilon}}}
				\AsymEq
				\left\{ \begin{aligned}
					1 - \frac{
						\epsilon^{2-\frac{d_S}{2}-1}
					}{
						B^\pm_l
					}, & \qquad d_S < 2, \\
					\tilde{\varphi}_l\sof{\pm\epsilon} 
					+
					\frac{
						B^\pm_l
					}{
						\brr{ u\sof{e^{i\Lambda^*_l}} }^2
					}
					\epsilon^{\frac{d_S}{2}-1}, & \qquad d_S > 2
				\end{aligned} \right.
				.
			\label{eq:AsymIntegrand}
			\end{equation}
			When $d_S$ is an even integer, logarithmic corrections appear.
			This case is discussed in Appendix \ref{app:Even}.
			An analytical remainder of $\varphi\sof{e^{i\lambda}}$ is always present (although it is trivial for small dimensions).
			Interestingly, we see that, in higher dimensions, additional constants appear.
			They are the derived from the return amplitudes:
			\begin{equation}
				u\sof{e^{i\Lambda^*_l}} 
				=
				\Sum{n=0}{\infty} 
				e^{in\Lambda^*_l} 
				u_n
				.
			\label{eq:DefAddConstants}
			\end{equation}
			These series converge for $d_S>2$.

			As we have mentioned, $u_n$ and $f\sof{E}$ as well as $\varphi_n$ and $\varphi\sof{e^{i\lambda}}$ are Fourier pairs.
			In the next section, we apply an asymptotic formula for Fourier transforms to relate Eq.~\eqref{eq:AsymIntegrand} with the large $n$ behavior of $\varphi_n$.
			After that, we apply the same formula to Eq.~\eqref{eq:SpecMeasAssump}.

	\section{Using a Fourier-Tauber formula}
	\label{sec:LargeN}
		According to Refs.~\cite{Erdelyi1956-0,Gamkrelidze1989-0}, the singular points of $\varphi\sof{e^{i\lambda}}$ and the power-law behavior around these points determine the large $n$ asymptotics of its Fourier coefficients, which are the first detection amplitudes $\varphi_n$.
		This is basically the Fourier analogue to the Tauberian theorems for the Laplace transform well-known in the theory of random walks \cite{Feller1971-0,Klafter2011-0}.
		The notable difference is that in the classical setup, there usually is only one singular point at vanishing Laplace variable with the consequence that the first passage probability decays monotonically in the long-time limit.
		This condition is violated in our case, as there are in general multiple singularities in $\varphi\sof{e^{i\lambda}}$.
		This fact is clearly related to the presence of quantum interference.
		Nevertheless, each of them can be isolated and the Tauberian theorem can be applied from both sides of the singular points.
		This leads to a sum of different power-law terms accompanied with a complex exponential factor in $n$. 
		A derivation of the formula is given in Appendix~\ref{app:Tauber}, while rigorous proofs are found in the above cited references.
		The main statement is the following:
		Let $h\sof{x}$ be a function with $L$ singularities at $x_l^*$, each admitting an expansion like Eq.~\eqref{eq:SpecMeasAssump}:
		\begin{equation}
			h\sof{x^*_l\pm\epsilon}
			\AsymEq
			\tilde{h}_l\sof{\pm\epsilon}
			+
			H^\pm_l \epsilon^{\nu-1}
			,
		\label{eq:HAssump}
		\end{equation}
		for some $l$-independent $\nu>0$.
		Then, its Fourier transform behaves for large $n$ like:
		\begin{equation}
			\frac{1}{2\pi}\Int{0}{2\pi}{x} e^{-inx} h\sof{x}
			\AsymEq
			\frac{\Gma{\nu}}{2\pi n^{\nu}} \Sum{l=0}{L-1} e^{-inx^*_l} \brr{
				\frac{H^+_l}{i^{\nu}}
				+
				\frac{H^-_l}{\sbr{-i}^{\nu}}
			}
			.
		\label{eq:HResult}
		\end{equation}
		In our prior publication \cite{Friedman2017-0,Friedman2017-1,Thiel2018-0} the large $n$ behavior was inferred from integrals along branch cuts in the complex plane.
		The same procedure would be viable here.
		In fact, each singular point $e^{i\Lambda^*_l}$ corresponds to a branch point of $\varphi\sof{z}$.
		Instead of writing $\varphi_n$ as a Fourier transform of $\varphi\sof{e^{i\lambda}}$ one could put the branch cuts of $\varphi\sof{z}$ along rays to complex infinity and integrate around them.
		However, the connection to the energy properties is clearer using the Fourier-Tauber theorem.

		Since $\varphi\sof{e^{i\lambda}}$ admits the expansion Eq.~\eqref{eq:AsymIntegrand} around each of the singular points $\Lambda^*_l$, one finds that $\varphi_n$ behaves for large $n$ like:
		\begin{equation}
			\varphi_n
			\AsymEq
			\Sum{l=0}{L-1}
			\frac{e^{-i n \Lambda^*_l}}{2\pi}
			\times
			\left\{ \begin{aligned}
				\tfrac{\sGma{2-\frac{d_S}{2}}}{n^{2-\frac{d_S}{2}}}
				\brr{
					\tfrac{i^{\frac{d_S}{2}}}{B^+_l}
					+
					\tfrac{\sbr{-i}^{\frac{d_S}{2}}}{B^-_l}
				}
				, & \; d_S < 2	\\
				\tfrac{\sGma{\frac{d_S}{2}}}{\sbrr{u\sof{e^{i\Lambda^*_l}}}^2 n^{\frac{d_S}{2}}}
				\brr{
					\tfrac{B^+_l}{i^{\frac{d_S}{2}}}
					+
					\tfrac{B^-_l}{\sbr{-i}^{\frac{d_S}{2}}}
				}
				, & \; d_S > 2	
			\end{aligned} \right.
		\label{eq:AsymFDA1}
		\end{equation}
		The squared absolute value of this expression is the desired first detection probability and reproduces Eq.~\eqref{eq:AsymFDP}.
		There, we hid most of the constants in the complex numbers $F_{l,d_S}$ which are now made explicit.
		Eq.~\eqref{eq:AsymFDP} [and Eq.~\eqref{eq:AsymFDA1}] is our main result and conveys the following qualitative properties:
		The first detection probability decays like a power law that only depends on the spectral dimension.
		The decay exponent exhibits a crossover at the critical dimension two and it is exactly double the exponent from the first passage probability of classical random walks \cite{Redner2007-0}.
		The frequencies of the oscillations are determined by the positions $E^*_l$ of the singularities of the MSDOS via Eq.~\eqref{eq:DefSingLambdas}.
		Eq.~\eqref{eq:AsymFDP} is valid for systems with a continuous energy spectrum.
		In the classical theory, the spectral dimension can always be identified with the exponent $d_S^\text{DOS}$ in $\rho\sof{E}$.
		It is hence a property of the Hamiltonian alone, independent of the initial or detection state.
		Such an identification is possible for ordinary detection states whose MSDOS $f\sof{E}$ behaves sufficiently smoothly around the van Hove singularities and also does not vanish at these points.
		This is basically a condition on the overlap of $\sKet{\PsiDet}$ with certain energy eigenstates, and has been used in a modified form also in \cite{Li2017-0}.

		A notable class of exceptions are those states that have no overlap with the eigenstates of the singular energies.
		We refer to them as ``insufficiently populated'' states.
		Consider the 1d tight-binding model with the detection state $\sKet{\PsiDet} = \sbr{\sKet{a} - \sKet{-a}}/\sqrt{2}$.
		Repeating the computations that led to Eq.~\eqref{eq:TBSMCounterEx1}, we find the MSDOS for this state to be:
		\begin{equation}
			f\sof{E}
			=
			\frac{1}{4\pi\gamma^2}\sqrt{E\sbr{4\gamma-E}}
			.
		\label{eq:TBSMCounterEx2}
		\end{equation}
		Although the density of states is an arcsin law, $f\sof{E}$ is a semicircle law.
		The spectral dimension in this case is $d_S = 3$ whereas $d_S^\text{DOS} = 1$!
		The reason is that $\sKet{\PsiDet}$ was chosen to have no momentum components with $k=0$ and $k=\pi/a$, which correspond to the singular energies.
		Hence, this $\sKet{\PsiDet}$ is insufficiently populated around the singular energies, leading to a discrepancy between $\rho\of{E}$'s and $f\sof{E}$'s spectral dimension.
		Another choice would be a state with wave vector representation supported only in the interval $[\pi/(2a),3\pi/(4a)]$.
		This example would even have different singular energies, because the support of $f\sof{E}$ lies in $[2\gamma,4\gamma\sin^2\sof{3\pi/8}]$.
		Yet another exception is a heavy-tailed detection state that decays like $\sBK{x}{\PsiDet} \AsymEq \sAbs{x}^{-1-\nu}$, for some $\nu \in [0,2]$.
		In momentum space, such a detection state will behave like $\sAbs{k}^\nu$ around the origin.
		This leads to a spectral dimension $d_S = 1+\nu$ as opposed to $d_S^\text{DOS} = 1$.
		All these cases show that the spectral dimension and the singular points, that we use in this article, are strictly speaking properties of $f\sof{E}$ and not of $\rho\sof{E}$.
		In many important cases, however, the positions of the singularities coincide in both, as do the power-law exponents.
		(Although the prefactors $A^\pm_l$ found in $f\sof{E}$ may be different from those found in $\rho\sof{E}$.)
		This is what we called ordinary.
		The classical theory does not know of insufficiently populated states.
		All eigenstates of the Hamiltonian/Laplacian are extended, due to the continuous nature of the spectrum, and have support in every lattice site.
		Hence, each lattice site has overlap with every energy, in particular with the singular ones.
		The only possible exception is a non-ergodic system that splits up into separate pieces.
		As a superposition of lattice states with negative interference in some energy is our of question due to positivity of probabilities, it is not possible to construct an insufficiently populated state in the classical first passage theory.

		The coefficients $B^\pm_l$ are often not easy to obtain.
		Knowledge about the return amplitudes $u_n$ opens up an alternate way to compute $\varphi_n$.
		We remember that $f\sof{E}$ and $u_n$ are also a Fourier pair and that we have access to $f\sof{E}$'s singularities in Eq.~\eqref{eq:SpecMeasAssump}.
		Therefore, the Fourier-Tauber theorem is now applied to Eq.~\eqref{eq:RA} using Eq.~\eqref{eq:SpecMeasAssump}.
		The result is:
		\begin{equation}
			u_n
			\AsymEq 
			\frac{1}{n^{\frac{d_S}{2}}}
			\Sum{l=0}{L-1} 
			C_l
			e^{-in\Lambda^*_l}
			,
		\label{eq:RAAsym}
		\end{equation}
		where 
		\begin{equation}
			C_l
			:=
			\Gma{\frac{d_S}{2}}
			\br{\frac{\hbar}{\tau}}^{\tfrac{d_S}{2}}
			\Sum{l'\sim l}{}
			\brr{
				\frac{
					A^+_{l'}
				}{
					i^{\frac{d_S}{2}}
				}
				+
				\frac{
					A^-_{l'}
				}{
					\sbr{-i}^{\frac{d_S}{2}}
				}
			}
			.
		\label{eq:DefC}
		\end{equation}
		In Eq.~\eqref{eq:RAAsym} we have already identified the critical angles $\Lambda^*_l = \tau E^*_l/\hbar$ from Eq.~\eqref{eq:DefSingLambdas}.
		The sum in Eq.~\eqref{eq:DefC} runs over all equivalent energies, i.e. $e^{i\tau E^*_{l'}/\hbar} = e^{i\Lambda^*_l}$.
		Remarkably, the return amplitudes oscillate with the same frequencies as do the detection amplitudes, which is also true in the above discussed exceptions, because $u_n$ and $\varphi_n$ are related to $f\sof{E}$ (and not to $\rho\sof{E}$).
		The relation between the MSDOS and the time decay of the return amplitudes is known in the literature \cite{Marchetti2012-0}.
		In Refs.~\cite{Muelken2006-0,Muelken2011-0} a similar argument was used to relate the spectral dimension to the decay of the return amplitudes.
		However, in these references the DOS was used instead of the MSDOS.
		Also the role of the van Hove singularities was under-appreciated.
		Therefore the authors were only able to predict the decay of the envelope of $u_n$ but not the oscillations.

		Often the matrix elements of the evolution operator are much more accessible than the MSDOS.
		Consequently the coefficients $C_l$ may be more easily available than the $B^\pm_l$'s.
		Therefore, we provide this alternative approach to derive $\varphi_n$.
		For example: In Eq.~\eqref{eq:RATB}, we found the return amplitudes for the tight-binding model in arbitrary dimension.
		Application of the asymptotic formula $\sBesselJ{0}{x} \AsymEq \cos\sof{x - \pi/4} \sqrt{2/(\pi x)}$, yields:
		\begin{align}
			u_n
			\AsymEq & \nonumber
			e^{-i\frac{2d\gamma\tau}{\hbar}n}
			\br{\frac{\hbar}{\pi \gamma\tau n}}^{\frac{d}{2}}
			\cos^d\of{\frac{2\gamma\tau}{\hbar}n - \frac{\pi}{4}}
			\\ = &
			\br{\frac{\hbar}{4\pi \gamma\tau n}}^{\frac{d}{2}}
			\Sum{l=0}{d}
			\binom{d}{l} 
			e^{-i\frac{4l\gamma\tau}{\hbar}n + i \sbr{2l-d}\frac{\pi}{4}}
			.
		\end{align}
		From here, one identifies:
		\begin{equation}
			C_l 
			=
			\br{\frac{\hbar}{4\pi\gamma\tau}}^{\frac{d}{2}}
			\binom{d}{l} 
			e^{i\sbr{2l-d}\frac{\pi}{4}}
			, \quad 
			\Lambda^*_l 
			= 
			\frac{4\gamma\tau}{\hbar} l
			,
		\label{eq:DefCTB}
		\end{equation}
		with $l\in\sbrrr{0,1,\hdots,d}$ for the tight-binding model.
		The above equation is correct, provided that $\tau$ does not assume a critical value.
		In that case, two singular points merge and the corresponding $C_l$'s have to be added.
		The points $4\gamma l$ are exactly the van Hove singularities of the density of states, see Fig.~\ref{fig:SM}(a,b).

		Multiplication of Eq.~\eqref{eq:RAAsym} with $r^n e^{i\lambda n}$, summing over $n$, and taking the limit $r\to1^-$, gives us the resolvent $u\sof{e^{i\lambda}}$.
		Close to the critical points $\Lambda^*_l$ it behaves as:
		\begin{align}
			u\of{e^{i\sbr{\Lambda^*_l\pm\epsilon}}}
			\AsymEq & \nonumber
			\tilde{u}_l\sof{\pm\epsilon}
			+
			\Gma{1-\tfrac{d_S}{2}}
			C_l
			\br{1-e^{\pm i\epsilon}}^{\frac{d_S}{2}-1}
			\\ \AsymEq &
			\tilde{u}_l\sof{\pm\epsilon}
			+
			\Gma{1-\tfrac{d_S}{2}}
			C_l
			\sbr{\mp i\epsilon}^{\frac{d_S}{2}-1}
			.
		\label{eq:}
		\end{align}
		The last line was obtained by taking $\sbr{1-e^{ix}}^\nu \AsymEq \sbr{-ix}^\nu$.
		Comparing this equation with \eqref{eq:AsymResolv} allows us to relate the different coefficients:
		\begin{equation}
			B^\pm_l
			= 
			\Gma{1-\tfrac{d_S}{2}} 
			e^{\mp i \tfrac{\pi\sbr{d_S-2}}{4}} 
			C_l 
			.
		\label{eq:Coeffs1}
		\end{equation}
		$B^\pm_l$ and $A^\pm_l$ can be related via Eq.~\eqref{eq:DefC}.
		Plugging the result into Eq.~\eqref{eq:AsymFDA1}, we can express the detection amplitudes in terms of the decay behavior of the return amplitudes (i.e. in terms of $C_l$):
		\begin{equation}
			\varphi_n
			\AsymEq
			\left\{ \begin{aligned}
				\frac{\br{1-\frac{d_S}{2}}\sin\of{\frac{\pi d_S}{2}}}{\pi n^{2-\frac{d_S}{2}}}
				\Sum{l=0}{L-1}
				\frac{e^{-i n \Lambda^*_l}}{C_l}
				, & \quad d_S < 2	\\
				\frac{1}{n^{\frac{d_S}{2}}}
				\Sum{l=0}{L-1}
				\frac{C_l}{\sbrr{u\sof{e^{i\Lambda^*_l}}}^2}
				e^{-i n \Lambda^*_l}
				, & \quad d_S > 2	
			\end{aligned} \right.
			.
		\label{eq:AsymFDA2}
		\end{equation}
		We used $\Gma{x}\Gma{1-x} = \pi/\sin\sof{x\pi}$.
		Alternatively, one could express the detection amplitudes in terms of the MSDOS $f\sof{E}$ (that is in terms of $A^\pm_l$) using Eq.~\eqref{eq:DefC}.
		For $d_S < 2$ one obtains: 
		\begin{equation}
			\varphi_n
			\AsymEq
			-\frac{1}{\pi^2n^2} \br{\frac{n\tau}{\hbar}}^{\frac{d_S}{2}}
			\Sum{l=0}{L-1}
			\frac{
				\Gma{2-\tfrac{d_S}{2}}
				\sin\of{\tfrac{\pi d_S}{2}}
				e^{-i n \Lambda^*_l}
			}{
				\sSum{l'\sim l}{}
				A^+_{l'} \sbr{-i}^{\frac{d_S}{2}}
				+ A^-_{l'} i^{\frac{d_S}{2}}
			}
			.
		\label{eq:AsymFDA3a}
		\end{equation}
		For $d_S > 2$ one obtains: 
		\begin{equation}
			\varphi_n
			\AsymEq
			\frac{\Gma{\frac{d_S}{2}}}{\br{\frac{n\tau}{\hbar}}^{\frac{d_S}{2}}}
			\Sum{l'=0}{L'-1} \frac{e^{-i n\frac{\tau E^*_{l'}}{\hbar}}}{u^2\sof{e^{i\frac{\tau E^*_{l'}}{\hbar}}}}
			\brr{
				A^+_{l'}\sbr{-i}^{\frac{d_S}{2}}
				+
				A^-_{l'} i^{\frac{d_S}{2}}
			}
			.
		\label{eq:AsymFDA3b}
		\end{equation}
		Eqs.~(\ref{eq:AsymFDA2}-\ref{eq:AsymFDA3b}) complement Eq.~\eqref{eq:AsymFDA1}.
		They are useful when neither the MSDOS $f\sof{E}$, nor an expansion of the return amplitudes, are available.

		Using Eq.~\eqref{eq:DefCTB} in Eq.~\eqref{eq:AsymFDA2} one obtains the detection amplitudes for the tight-binding model for dimensions larger than $2$:
		\begin{equation}
			\varphi_n
			\AsymEq
			\br{\frac{\hbar}{4\pi \gamma\tau n}}^{\frac{d}{2}}
			\Sum{l=0}{d}
			\binom{d}{l}
			\frac{
				e^{-i\frac{4l\gamma\tau}{\hbar} n + i \sbr{2l-d}\tfrac{\pi}{4}}
			}{
				\sbrr{u\sof{e^{i\frac{4l\gamma\tau}{\hbar}}}}^2
			}
			.
		\label{eq:TBFDA}
		\end{equation}
		The modulus squared of this expression is the first detection probability:
		\begin{equation}
			F_n
			\AsymEq
			\br{\frac{\hbar}{4\pi \gamma\tau n}}^{d}
			\Abs{
				\Sum{l=0}{d}
				\binom{d}{l}
				\frac{
					e^{i\sbr{d-2l}\br{\frac{2\gamma\tau}{\hbar} n -\tfrac{\pi}{4}} - 2i\arg{u\sof{e^{i\frac{4l\gamma\tau}{\hbar}}}}}
				}{
					\sAbs{u\sof{e^{i\frac{4l\gamma\tau}{\hbar}}}}^2
				}
					}^2
			.
		\label{eq:TBFDPHighD}
		\end{equation}
		From the expression of $u_n$, Eq.~\eqref{eq:RATB}, and the definition of $u\sof{z}$, one can infer that $u\sof{e^{i4l\gamma\tau/\hbar}}$ is the complex conjugate of $u\sof{e^{i4\sbr{d-l}\gamma\tau/\hbar}}$.
		Hence, for each term in the sum of Eq.~\eqref{eq:TBFDPHighD}, its complex conjugate appears as well.
		The complex exponentials can actually be replaced by cosines and half of the terms in the sum can be dropped.
		The constants $u\sof{e^{i4l\gamma\tau/\hbar}}$ are defined by Eq.~\eqref{eq:DefAddConstants}, no closed form is known to us, and so they have to be computed numerically.

		Eq.~\eqref{eq:TBFDPHighD} holds for all dimensions larger than two.
		The one dimensional tight-binding model has already been discussed extensively in Refs.~\cite{Friedman2017-0,Friedman2017-1,Thiel2018-0}.
		There, the following formula was reported:
		\begin{equation}
			F_n
			\AsymEq
			\frac{4\gamma\tau}{\hbar \pi n^3}
			\cos^2\of{\frac{2\gamma\tau}{\hbar} n + \frac{\pi}{4}}
			,
		\label{eq:TBFDP1D}
		\end{equation}
		which is perfect accordance with Eq.~\eqref{eq:AsymFDA2}.
		Curiously, in both the one- and three dimensional case, one finds a power law with exponent $-3$.

		In the two dimensional case logarithmic corrections to the power law appear.
		This case is discussed in Appendix~\ref{app:Even}.
		We simply state here the result for the 2d tight-binding model:
		\begin{equation}
			F_n
			\AsymEq
			\br{ \frac{ 4\pi\gamma\tau}{\hbar n \ln^2 n}}^2
			\Abs{
				\frac{1}{2}
				-
				2\sin\of{\frac{4\gamma\tau}{\hbar} n}
			}^2
			.
		\label{eq:TBFDP2D}
		\end{equation}

		In Eq.~\eqref{eq:TBFDP1D} and Eq.~\eqref{eq:TBFDP2D} the Zeno effect is visible.
		When the limit $\tau\to0$ is taken, our asymptotic result vanishes.
		From Eq.~\eqref{eq:QuantumRenewal} and Eq.~\eqref{eq:RATB}, one finds that actually $F_n \to \delta_{n,1}$.
		The physical meaning is that the particle is found at the detection site, immediately after the experiment commences.
		Hence the long-time asymptotics of $F_n$ vanish.
		The expression in Eq.~\eqref{eq:TBFDPHighD}, however, diverges for fixed $n$ in the limit $\tau\to0$!
		(This can be seen from numerical evaluation of $u\sof{e^{i4l\gamma\tau/\hbar}}$.)
		In the numerical simulations of $F_n$, an intermediary regime appears in $F_n$ that is not described by our asymptotic formula.
		As $\tau$ decreases, this intermediary region grows in size.
		The $F_n$ values in that regime go to zero as $\tau$ decreases.
		At the same time, $F_1$ goes to unity, so that in total $F_n\to \delta_{n,1}$, and the Zeno effect is restored.
		It can not be observed in our asymptotic formula, though.

\begin{figure*}
			\includegraphics[width=0.99\textwidth]{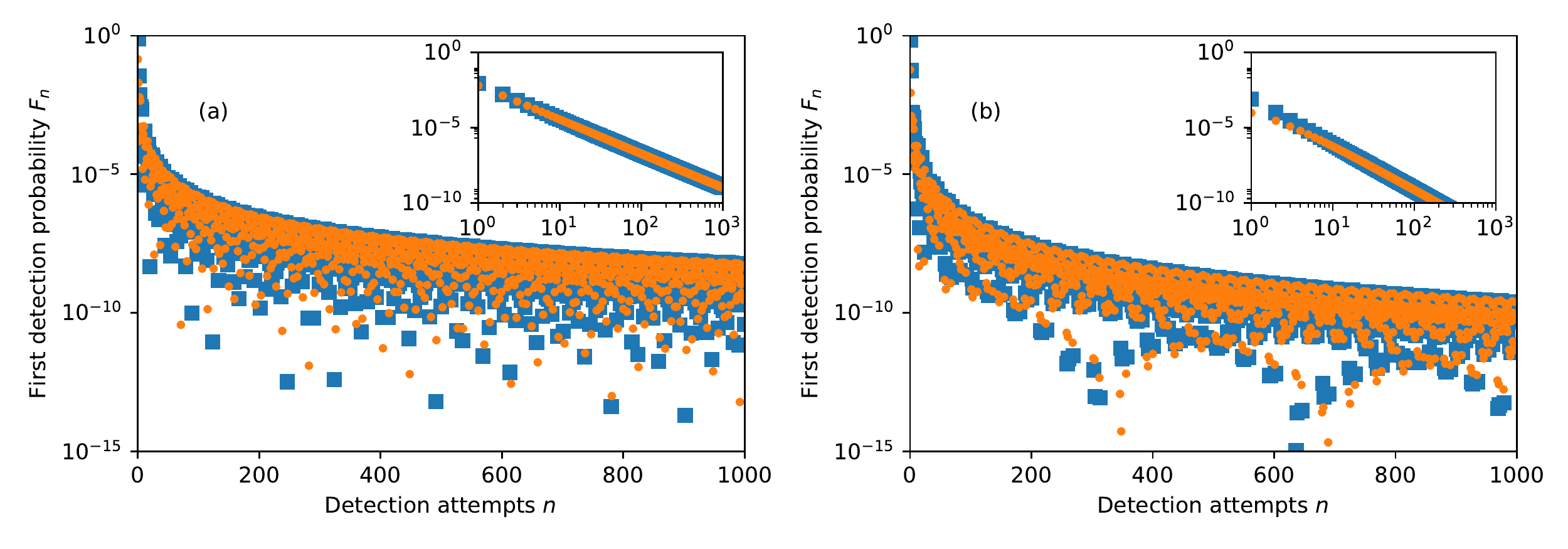}
			\caption{
				Probability of first detected return for the tight-binding model in a semi-logarithmic plot.
				Blues squares are numerical results, orange circles are the predictions of Eq.~\eqref{eq:TBFDP2D} and Eq.~\eqref{eq:TBFDPHighD}.
				(a): Two dimensional case with $\tau = 0.25\hbar/\gamma$.
				The $n^{-2}$ power-law and the oscillations are clearly visible.
				The logarithmic factor was adjusted to $(\pi + \ln n)^{-4}$, see Appendix~\ref{app:Even}.
				Inset: Two dimensional case for the critical detection period $\tau = \tau_c = (\pi/2)\hbar/\gamma$.
				No oscillations are present here, because $\mu\sof{\lambda}$ has only one singular point.
				We find a $n^{-2}$ power law with logarithmic factor $(x + \ln n)^{-4}$, where $x \ApproxEq 10.531$, see Appendix~\ref{app:Even}.
				(b): Three dimensional case for $\tau = 0.25 \hbar/\gamma$.
				Inset: $d=3$ with $\tau = \tau_c = \sbr{\pi/2}\hbar/\gamma$.
				Oscillations are only present for non-critical $\tau$.
				No fitting was applied to the right-hand side plots.
				\label{fig:TB_FDP}
			}
		\end{figure*}

		Numerical simulations were performed by using Eq.~\eqref{eq:RATB} in Eq.~\eqref{eq:QuantumRenewal} and solving the resulting system of linear equations for $\varphi_n$.
		In Fig.~\ref{fig:TB_FDP}, we present numerical simulations for the tight-binding model in dimensions two and three for $\tau = 0.25\hbar/\gamma$.
		In the two-dimensional case, we fitted the envelope of the asymptotic result \eqref{eq:TBFDP2D}.
		Due to the slow logarithmic convergence it is necessary to replace $\ln^2n$ by $\sbr{\ln n + x}^2$.
		$x$ is determined from a fit.
		Both expressions are equivalent in the asymptotic limit.
		The figure depicts the fit.

		One clearly sees the expected dimension dependent power-law decay of Eq.~\eqref{eq:AsymFDP} in the envelope of $F_n$ as well as the oscillations, and an overall good agreement with our prediction.

\begin{figure}
			\includegraphics[width=0.99\columnwidth]{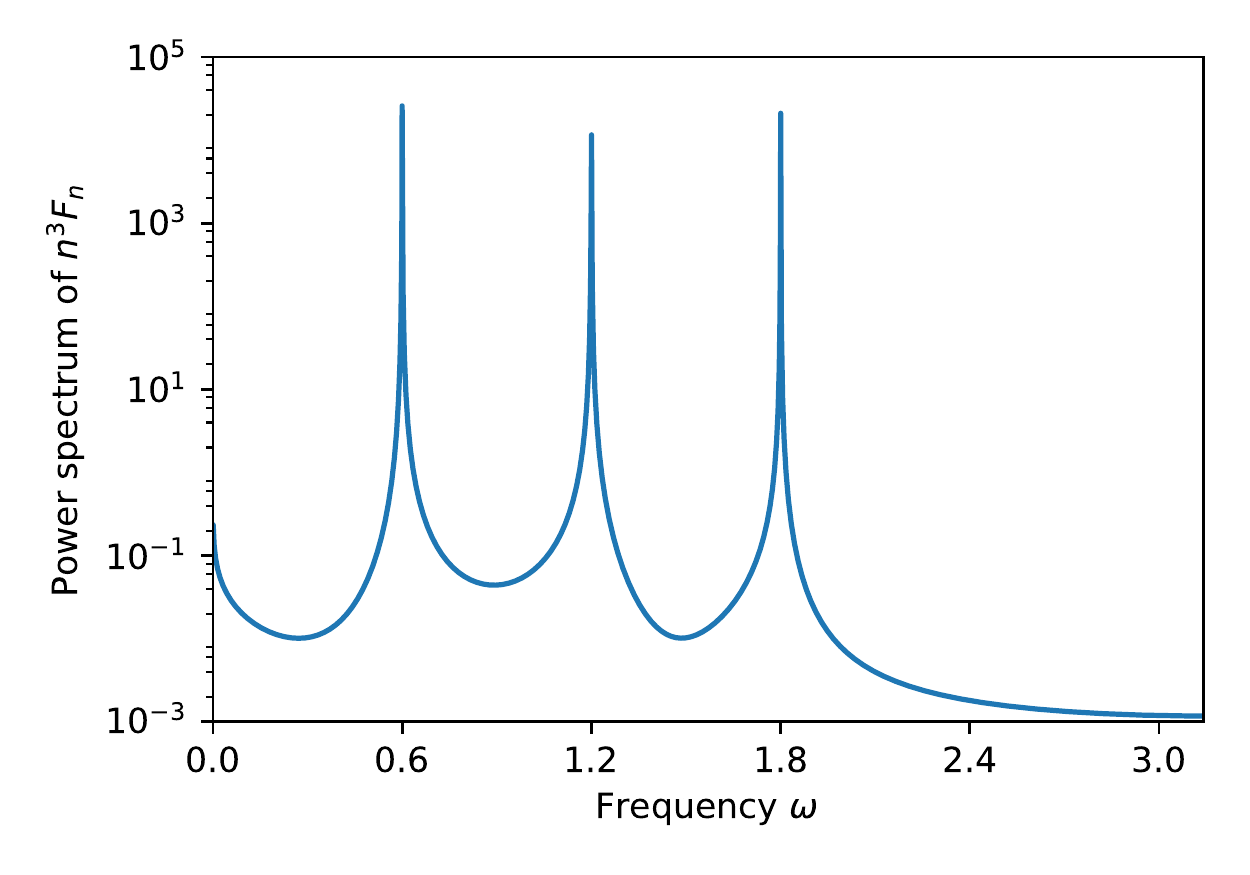}
			\caption{
				Power spectrum of $n^3 F_n$ for $d=3$ and $\tau=0.15\hbar/\gamma$.
				Stripping off the power law off of $F_n$ by the multiplication with $n^3$ exposes the oscillatory terms.
				The frequencies of the oscillations correspond to peaks in the Fourier transform.
				Compare the position of the peaks with the prediction of equation Eq.~\eqref{eq:DefSingLambdas}, $\Lambda^*_l = 0.6 l$, $l=0, 1, 2, 3$.
				\label{fig:TB_FFT}
			}
		\end{figure}

		As discussed, the frequency of the oscillations is given by Eq.~\eqref{eq:DefSingLambdas} and can be controlled via the detection period $\tau$.
		This is demonstrated in Fig.~\ref{fig:TB_FFT}, where we plotted the power spectrum of $n^3 F_n$ for $d=3$ and $\tau =0.15\hbar/\gamma$.
		(The power spectrum is the modulus squared of the discrete Fourier transform of $n^3 F_n$.)
		This makes it possible to visualize the characteristic frequencies.
		The frequencies $\Lambda^*_l = 0.6 l$, $l\in\sbrrr{0,1,2,3}$ are visible as peaks in the power spectrum of $n^3 F_n$.
		As another demonstration, we present the case of a critical detection amplitude, $\tau_c = (\pi/2) \hbar/\gamma$ from Eq.~\eqref{eq:TBDefCritTau}.
		For this value of $\tau$, all critical energies become equivalent and get mapped to the same critical point $\Lambda^*$.
		As a consequence the oscillations in $F_n$ disappear and only the power law remains.
		This is shown in the insets of Fig.~\ref{fig:TB_FDP}.

\begin{figure*}
			\includegraphics[width=0.99\textwidth]{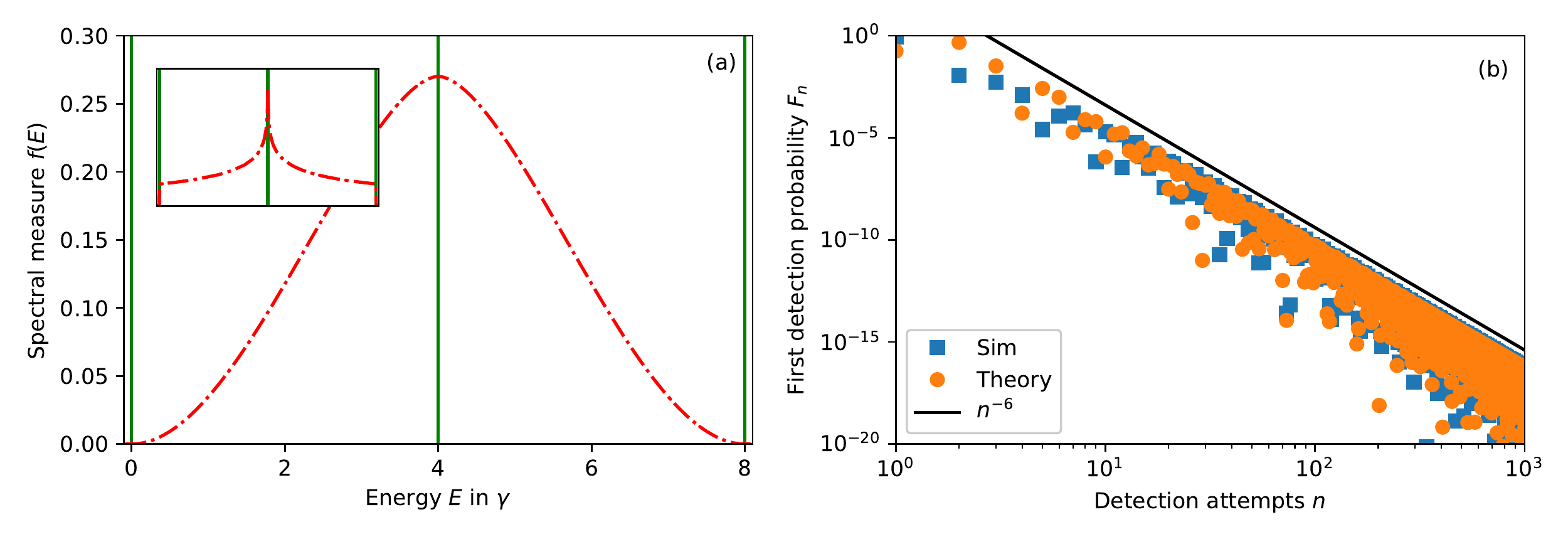} 
			\caption{
				Two dimensional tight-binding model with the special detection state of Eq.~\eqref{eq:PsiDetSpecial}.
				(a) The MSDOS $f\sof{E}$.
				It is wildly different from the DOS $\rho\sof{E}$ depicted in the inset or in detail in Fig.~\ref{fig:SM}(a).
				The spectral dimension found in $f\sof{E}$ is six rather than two, which is obtained from the DOS.
				The singularities of $f\sof{E}$ are in the in its second derivative, but they can be found in the same positions as the singularities of $\rho\sof{E}$.
				(b) The first detection probabilities $F_n$ for this case.
				In contrast to a ordinary choice of the detection state, we do not find a $n^{-2}\ln^{-4} n$ decay of $F_n$ but rather a $n^{-6}$ decay in accordance with Eq.~\eqref{eq:AsymFDP} for large spectral dimensions.
				\label{fig:TBSpecial}
			}
		\end{figure*}

		Finally, we want to illustrate our discussion on insufficiently populated states and we want to showcase the dependence of the spectral dimension on the detection state.
		Therefore, we consider the two dimensional tight-binding model with the detection state:
		\begin{equation}
			\sKet{\PsiDet}
			=
			\frac{1}{2} \brr{ 
				\sKet{\sbr{a,a}}
				+
				\sKet{\sbr{-a,-a}}
				-
				\sKet{\sbr{a,-a}}
				-
				\sKet{\sbr{-a,a}}
			}
			.
		\label{eq:PsiDetSpecial}
		\end{equation}
		By construction it has no overlap with the singular energies $E^*_l = 0, 4\gamma, 8\gamma$.
		The corresponding MSDOS is a convolution of Eq.~\eqref{eq:TBSMCounterEx2} with itself.
		This is a convolution of two MSDOS's with $d_S = 3$ each, resulting in $d_S = 6$.
		$f\sof{E}$ is presented in Fig.~\ref{fig:TBSpecial}(a) and does not resemble Fig.~\ref{fig:SM}(a) at all.
		For this choice, the return amplitudes are given by:
		\begin{equation}
			u_n
			=
			e^{-i\frac{4\gamma\tau}{\hbar}n}
			\brr{ 
				\BesselJ{0}{\frac{2\gamma\tau}{\hbar}}
				+
				\BesselJ{2}{\frac{2\gamma\tau}{\hbar}}
			}^2
		\label{eq:}
		\end{equation}
		and decay like $n^{-3}$.
		The spectral dimension in $f\sof{E}$ is $d_S = 6$ instead of $d_S^\text{DOS}=2$ found in the DOS!
		The simulations depicted in Fig.~\ref{fig:TBSpecial}(b) reflect this fact nicely.

		In the next two sections we discuss two off-lattice models.
		The first is the free particle in continuous space and the second is the L\'evy particle.

	\section{The free particle in continuous space}
	\label{sec:FreePart}
		\begin{figure*}
			\includegraphics[width=0.99\textwidth]{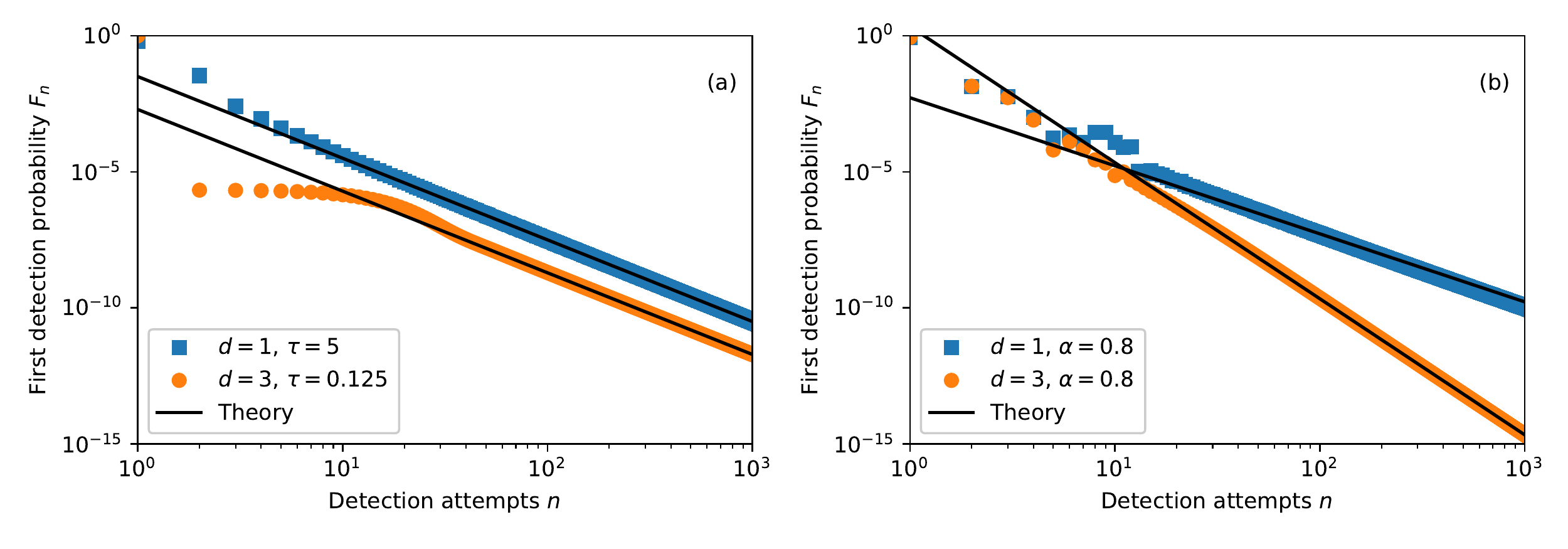}
			\caption{
				First detection probabilities for the free particle and the L\'evy particle.
				Blue squares: $d=1$, Orange circles: $d=3$, Black lines: theoretical prediction of Eq.~\eqref{eq:NRFDP} and Eq.~\eqref{eq:LevyFDP}
				(a): Free particle. $\tau$ is measured in units of $M\sigma^2/\hbar$.
				As $\tau$ decreases, $F_1$ moves closer to unity and a plateau appears for small $n$.
				The plateau extends to the right as $\tau$ becomes smaller and the plateau value converges to zero.
				This is the Zeno effect $F_n\to\delta_{n,1}$.
				Our asymptotic solution only describes the non-plateau regime of $F_n$.
				(b) L\'evy particle with $\tau = 1 \hbar/E_0$, $\alpha=0.8$ and different $d$.
				Prediction and simulations agree nicely.
				By tuning $\alpha$ arbitrary power law exponents can be observed.
				\label{fig:FreePart}
			}
		\end{figure*}
		We now consider a free non-relativistic particle with mass $M$ in continuous $d$-dimensional space.
		The Hamiltonian is given by the kinetic energy
		\begin{equation}
			\Ham := \frac{\hbar^2}{2M} \hat{\V{k}}^2
			,
		\label{eq:FPHamiltonian}
		\end{equation}
		where we write its momentum in terms of the wave vector $\hbar \V{k}$.
		The first difficulty one faces here is the definition of the detection state.
		In contrast to the lattice system, position eigenstates have zero width, and projection onto such states can be tricky.
		We assume instead that our detector has a finite accuracy $\sigma$ and projects the wave function to a Heisenberg state with minimum uncertainty.
		In momentum representation this state is defined as:
		\begin{equation}
			\psi_{\V{0}}\of{\V{k}}
			:=
			\br{\frac{2\sigma^2}{\pi}}^{\frac{d}{4}}
			\exp\off{-\sigma^2\V{k}^2}
			,
		\label{eq:NRDefPsi}
		\end{equation}
		where the wave vector/momentum $\V{k} = \V{p}/\hbar$ is a real $d$-dimensional vector.
		Such a wave function has $\sEA{\V{r}} = \V{0}$, $\sEA{\V{p}}=\V{0}$, furthermore: $\sEA{\sbr{\V{r}-\V{x}_0}^2} = d \sigma^2$, and $\sEA{\V{p}^2} = d \hbar^2/(4 \sigma^2)$.
		In particular it has minimum uncertainty between the momentum and position coordinates.
		We choose this state as initial and detection state: $\sKet{\PsiDet} = \sKet{\PsiIn} = \sKet{\psi_{\V{0}}}$.

		For the free particle the dispersion relation is the usual kinetic energy:
		\begin{equation}
			E\of{\V{k}}
			=
			\frac{\sbr{\hbar\V{k}}^2}{2M}
			.
		\label{eq:NRDispersion}
		\end{equation}
		As an abbreviation, we define the detection state's energy per degree of freedom: $E_0 := \EA{\V{p}^2/(2m)}/(2d) = \hbar^2/(4m\sigma^2)$.
		Using the momentum representation of the evolution operator, we can obtain the return amplitudes from an integral over all wave vectors:
		\begin{align}
			\sBAK{\psi_{\V{0}}}{\TEO\of{n\tau}}{\psi_{\V{0}}}
			= &
			\Int{\Reals^d}{}{\V{k}}
			\sAbs{\psi_{\V{0}}\sof{\V{k}}}^2 e^{-i \frac{n\tau E\sof{\V{k}}}{\hbar}}
			=
			\frac{1}{\br{1 + i \frac{E_0 \tau}{\hbar}n}^{\frac{d}{2}}}
			\label{eq:NRTEO}
			.
		\end{align}
		The expression has two regimes for $n$: 
		When $1 \gg n E_0\tau/\hbar = (n \hbar \tau) /(4M\sigma^2) $, the return amplitude is approximately unity.
		For fixed $n$, this is the case when either $\tau$ is very small or $\sigma$ is very large.
		In the opposite case, the return amplitude decays like a power law, which reveals the spectral dimension as equal to the Euclidean one,$d_S =d$.
		The crossover between the two regimes appears at $\tilde{n} \approx \hbar/(E_0\tau) = 4M\sigma^2/(\hbar\tau)$ and marks the time at which the momentum-induced dispersion of the wave packet becomes comparable to the initial width of the wave packet.
		In the limit $\sigma\to0$, the return amplitudes vanish.
		This signals to us that an infinitely precise position measurement is not physically meaningful.

		Since the dispersion relation \eqref{eq:NRDispersion} is a parabola, there is only one critical point on the energy surface, which is located at zero momentum.
		This shows there will be no oscillations in the first detection probabilities in accord with the simulations depicted in Fig.~\ref{fig:FreePart}(a).
		The resolvent at the critical value on the unit circle can be expressed via the Hurwitz zeta function $\zeta\sof{s;a} := \sSum{n=0}{\infty} \sbr{n+a}^{-s}$.
		\begin{equation}
			u\sof{z=1}
			=
			\br{\frac{-i\hbar}{E_0\tau}}^{\frac{d}{2}}
			\zeta\sof{\tfrac{d}{2};-i\tfrac{\hbar}{E_0\tau}}
			,
		\label{eq:NRResolv}
		\end{equation}

		From the momentum presentation and the dispersion relation it is also easy to find the MSDOS.
		\begin{equation}
			f\sof{E}
			=
			\Int{\Reals^d}{}{\V{k}} \sAbs{\psi_{\V{0}}\sof{\V{k}}}^2 
			\delta\sof{E - \tfrac{\sbr{\hbar \V{k}}^2}{2m}}
			=
			\frac{e^{-\frac{E}{E_0}}}{\Gma{\tfrac{d}{2}} E_0}
			\br{\tfrac{E}{E_0}}^{\frac{d}{2}-1}
			.
		\label{eq:NRSpecMeas}
		\end{equation}
		Of course, $f\sof{E} = 0$ for $E<0$.
		Therefore, one can easily identify the constants $A^\pm$ of the MSDOS's singularity at $E^*=0$:
		\begin{equation}
			A^+ := \frac{1}{\Gma{\tfrac{d}{2}} E_0^{\frac{d}{2}} }, \quad A^- = 0, \quad d_S = d
		\label{eq:NRA}
		\end{equation}
		Using Eq.~\eqref{eq:NRResolv} and \eqref{eq:NRA} in Eqs.~\eqref{eq:AsymFDA3a} and \eqref{eq:AsymFDA3b}, and squaring the result, one obtains:
		\begin{equation}
			F_n
			\AsymEq
			\left\{ \begin{aligned}
				\frac{E_0\tau}{4 \pi^2 \hbar n^3}
				, & \quad d = 1	\\
				\br{\frac{E_0\tau}{\hbar n \ln^2 n}}^2
				, & \quad d = 2 \\
				\Abs{\zeta\sof{\tfrac{d}{2};-i\tfrac{\hbar}{E_0\tau}}}^{-4}
				\br{\frac{E_0 \tau}{\hbar n}}^{d}
				, & \quad d > 2 	
			\end{aligned} \right.
			.
		\label{eq:NRFDP}
		\end{equation}
		(For the 2d case see Appendix~\ref{app:Even}.)
		Fig.~\ref{fig:FreePart} shows excellent agreement between simulations and Eq.~\eqref{eq:NRFDP}.
		To our surprise, the Zeno effect is not visible in our equation for $d>3$!
		Using the formula $\sAbs{\zeta\sof{s;-i/x}} \AsymEq x^{s-1}$, as $x\to0$, we find that $F_n \PropTo \tau^{4-d}$.
		In our simulations, however, we still find $F_n\to\delta_{n,1}$ as $\tau\to0$.
		As $\tau$ decreases a plateau forms in $F_n$ for small $n$ that increases in size and decreases in height.
		In the same time $F_1$ moves closer to unity.
		The asymptotic formula is valid only in the non-plateau region.
		This is similar to the tight-binding case.

	\section{Free L\'evy-particle}
	\label{sec:Levy}
		Instead of the regular dispersion relation \eqref{eq:NRDispersion}, one can also impose an anomalous energy-momentum relation:
		\begin{equation}
			E\sof{\V{k}}
			=
			C \sAbs{\V{k}}^\alpha
			,
		\label{eq:LevyDisp}
		\end{equation}
		with $0<\alpha<2$, and some constant $C$ with suitable units.
		This can be viewed as a continuous interpolation between a non-relativistic and a relativistic dispersion relation, the latter being attained by putting $\alpha=1$.
		Such a dispersion relation can be obtained, when the Laplacian in the Hamiltonian is replaced by a fractional Laplacian (a Riesz-Feller derivative) of order $\alpha/2$, which is obviously very different than the canonical approach that leads to the Dirac equation.

		Detection and preparation state are again taken to be the Gaussian, Eq.~\eqref{eq:NRDefPsi}.
		We define the energy constant $E_0 := C (2\sigma^2)^{-\alpha/2}$ which is related to the energy of the state $\sKet{\psi_{\V{0}}}$.
		Using Eq.~\eqref{eq:NRDefPsi} and the dispersion relation, we can write the return amplitude in terms of an integral.
		We use spherical coordinates and change variables to $y = 2\sigma^2 k^2$, to obtain:
		\begin{equation}
			\sBAK{\psi_{\V{0}}}{\TEO\sof{t}}{\psi_{\V{0}}}
			=
			\frac{1}{\Gma{\frac{d}{2}}}
			\Int{0}{\infty}{y}
			y^{\frac{d}{2}-1}
			e^{-y - i \frac{E_0 t}{\hbar} y^{\frac{\alpha}{2}}}
			.
		\label{eq:LevyTEO}
		\end{equation}
		For large $t$, expanding the $e^{-y}$ term leads to an asymptotic series in inverse powers of $t$.
		The leading order is $t^{-d/\alpha}$.
		Therefore the spectral dimension is $d_S = 2d/\alpha$ and can be tuned to any real number larger than $d$ by adjusting $\alpha$.
		The same result is obtained when computing $f\sof{E}$ from the Brillouin zone:
		\begin{equation}
			f\sof{E}
			=
			\frac{2}{\alpha}\frac{e^{-\br{\frac{E}{E_0}}^{\frac{2}{\alpha}}}}{\Gma{\tfrac{d}{2}}E_0}
			\br{\frac{E}{E_0}}^{\frac{d}{\alpha}-1}
			.
		\label{eq:LevySpecMeas}
		\end{equation}
		From here we identify the coefficients $A^\pm$:
		\begin{equation}
			A^+ 
			:=
			\frac{2}{\alpha}\frac{1}{\Gma{\tfrac{d}{2}}}
			E_0^{-\frac{d}{\alpha}}
			, \quad A^- = 0
			, \quad d_S = \frac{2}{\alpha} d,
		\label{eq:LevyA}
		\end{equation}
		and write down the first detection probabilities:
		\begin{equation}
			F_n
			\AsymEq
			\left\{ \begin{aligned}
				\frac{\br{1-\frac{d}{\alpha}}^2}{\pi^2 n^{4 - \frac{2d}{\alpha}}}
				\br{\frac{\Gma{\frac{d}{2}+1}}{\Gma{\frac{d}{\alpha}+1}}}^2
				\br{ \frac{\tau E_0}{\hbar}}^{\frac{2d}{\alpha}}
				, & \quad d < \alpha	\\
				\br{\frac{\tau E_0\Gma{1+\tfrac{d}{2}}}{\hbar n \ln^2 n} }^2
				, & \quad d = \alpha \\
				\br{\frac{\Gma{\frac{d}{\alpha}+1}}{\Gma{\frac{d}{2}+1}}}^2
				\frac{1}{\sAbs{u\sof{z=1}}^4}
				\br{ \frac{\hbar}{n \tau E_0} }^{\frac{2d}{\alpha}}
				, & \quad d > \alpha 	
			\end{aligned} \right.
			.
		\label{eq:LevyFDP}
		\end{equation}
		The remaining constant can be computed via a numeric integral:
		\begin{equation}
			u\sof{z=1}
			=
			\frac{1}{\Gma{\frac{d}{2}}}
			\Int{0}{\infty}{y}
			\frac{y^{\frac{d}{2}-1} e^{-y}}{
				1 - e^{-i\frac{E_0\tau}{\hbar} y^{\frac{\alpha}{2}}}
			}
		\label{eq:}
		\end{equation}
		Simulations of such a process are possible by numerically evaluating the integral of Eq.~\eqref{eq:LevyTEO}.
		Results are depicted in Fig.~\ref{fig:FreePart}(b).
		This simple but important example shows that Eq.~\eqref{eq:AsymFDP} also holds for {\em fractional} values of the spectral dimension.

	\section{Discussion}
		\label{sec:Disc}
		This article exposed the relation between the quantum first detection probability and the MSDOS $f\sof{E}$.
		Qualitative features like the power law decay and the frequencies of the oscillations have been obtained from $f\sof{E}$, in particular from its singular points and its behavior in their vicinity.
		In the ordinary case these properties can also be inferred from the DOS $\rho\sof{E}$.

		We stress that our main results from Eq.~\eqref{eq:AsymFDP}, also hold for the problem of the first detected {\em arrival}, i.e. when $\sKet{\PsiIn} \ne \sKet{\PsiDet}$.
		The additional steps are carried out in Appendix~\ref{app:Arrival}.
		In the main text, we only restricted ourselves to the problem of first detected return for notational economy.

		The frequencies of the oscillations in $F_n$ can be found in the asymptotic decay of the return amplitudes.
		This is clear from comparison of Eq.~\eqref{eq:RAAsym} and Eq.~\eqref{eq:AsymFDA1}.
		The reason is that they both are related via $f\sof{E}$.
		The $\tau$-dependence of the frequencies gives rise to the existence of critical detection periods, when the number of different frequencies changes abruptly.
		In the ordinary case, the frequencies are determined by the van Hove singularities.
		
		The power law exponents of Eq.~\eqref{eq:AsymFDP} are the exact double of the classical exponents of the first passage problem for random walks.
		The hand-waving argument is that Eq.~\eqref{eq:QuantumRenewal} is an equation for amplitudes, whereas its classical analogue Eq.~\eqref{eq:ClassicalRenewal} is an equation for probabilities.
		The necessary squaring operation brings the additional factor of two in the exponent.
		This exponent doubling when going from the classical to the quantum problem was also reported in Refs.~\cite{Muelken2006-0,Muelken2011-0} for the return probability, and also in Ref.~\cite{Boettcher2015-0} where it manifested in a halving of the walk dimension for certain discrete-time quantum walks.
		However, only for ordinary states the relevant spectral dimension $d_S$ agrees with the one found in the DOS, $d_S^\text{DOS}$.
		
		As a consequence of the large decay exponents of $F_n$, it assumes substantial values only for small $n$.
		The larger exponent comes with the price that the quantum system is not {\em almost surely detectable}, in the sense that the total probability of detection, $P_\text{det} = \sSum{n=1}{\infty} F_n$, is always smaller than unity.
		This holds unless $\mu\sof{\lambda}$ is a sum of delta functions \cite{Gruenbaum2013-0}.
		Hence, in our infinite space models, featuring a continuous energy spectrum, there is always a non-zero probability that the particle escapes the detector.
		Such a deficit in the total return probability is also known in the classical problem, where it marks the dichotomy between recurrent and transient random walks \cite{Polya1921-0}.
		Random walks with a spectral dimension smaller than the critical dimension two will eventually return to their initial position with probability one and therefore are considered recurrent.
		For transient random walks with a spectral dimension larger than two there is a finite probability that they never return to their initial site.

		Although the total detection probability is smaller than unity, one can compute the average first detection time: $\sEA{n} = \sSum{n=1}{\infty} F_n n / P_\text{det}$ {\em under the condition} that the system was detected at all.
		As a consequence of the larger exponents we found (the slowest decay is $n^{-2}\ln^{-4}n$), this expectation is always finite, contrasting the classical situation.
		The conditional variance of the first detection time is finite only in dimensions larger than three.

		We found that $d_S = 2$ is a critical dimension, which also plays an important role for quantum search algorithms.
		For a coined quantum search algorithm in dimensions larger than two, the Grover efficiency $\sLandau{\sqrt{N}}$ can be attained \cite{Aaronson2003-0}.
		In coinless, oracle quantum searches, the critical dimension is four \cite{Childs2004-0,Li2017-0}.
		However, the last reference shows that the decisive quantity is exactly the {\em spectral} dimension, just as in our problem.

		Finally, our main assumption for the identification of the spectral dimensions found in the DOS and in the MSDOS, is that neither $\sKet{\PsiDet}$ nor $\sKet{\PsiIn}$ are insufficiently populated, i.e. they both overlap with the critical van Hove energies.
		Exceptions have been discussed in the main text and show that the first detection probabilities subtly depend on the initial and detection state.
		This is in sharp contrast to the classical problem, where the particular choice of initial state often only changes the transient, but not the long-time behavior of the first passage probability.
		This new dependence on the detection state is encoded in the MSDOS $f\sof{E}$ associated with the detection state.
		As we have shown in the main text this important quantity is related to, but ultimately different from the DOS.
		It requires a high degree of symmetry in the system and a special choice of states for the DOS and the MSDOS to coincide.
		We felt that $f\sof{E}$ is rarely known outside the mathematical community, hence we focused on ``ordinary'' situations where the spectral dimension and the singularities of $f\sof{E}$ can also be found in the well-known density of states.
		Still, this state-dependence can lead to very counter-intuitive results, if one does not carefully confirm the overlap condition.

		Many open questions remain for the future.
		These include the details of the arrival problem, for instance the dependence of the amplitudes $F_{l,d_S}$ from Eq.~\eqref{eq:AsymFDP} on the distance between initial an detector position.
		We do not have a clear intuition on the situation when the initial state is insufficiently populated, but the detection state is not.
		Also unclear is what happens when the energy spectrum is neither completely discrete nor completely continuous, that is when it is singular continuous as in the Aubrey-Harper model \cite{Lahiri2017-0}.
		It is possible to extend our arguments to coined quantum walks, by discussing the WMSDOS $\mu\sof{\lambda}$ of their evolution operator.

		Our most surprising finding is the sensitivity to the initial and detection states for out-of-the-ordinary states, which is a purely quantum phenomenon related to interference.
		This necessitates an adjustment of the concept of spectral dimension from $d_S^\text{DOS}$, defined by the DOS, to $d_S$ defined by the MSDOS.

		We are confident, that this work will lift the quantum first detection theory closer to the level of its classical brother, the first passage theory of random walks.

	\acknowledgements
		The authors acknowledge support from the Israeli Science Foundation under Grant No. 1898/17.
		FT is funded by the Deutsche Forschungsgemeinschaft under grant TH-2192/1-1.

	\appendix
	\section{Behavior of the resolvent close to the unit circle}
	\label{app:Plemelj}
		In this appendix, we derive Eq.~\eqref{eq:AsymResolv} of the main text.
		The main part of this derivation is concerned with an analogue of the classic formula $\sbr{x-i\epsilon}^{-1} \to P.V.\; 1/x +  i\pi \delta\sof{x}$ for the unit circle.
		This shows that $u\sof{re^{i\lambda}}$ is related to the WMSDOS and its Hilbert transform:
		\begin{equation}
			u\sof{re^{i\lambda}}
			\underset{r\to1^-}{\longrightarrow}
			\frac{
				1
				+
				\mu\of{\lambda}
				+ i \Hilbert{\mu}\of{\lambda} 
			}{2}
			.
		\label{eq:Plemelj}
		\end{equation}
		The Hilbert transform $\Hilbert{\mu}$ on the unit circle is defined by a principal value integral: 
		\begin{equation}
			\Hilbert{\mu}\of{\lambda}
			:=
			\frac{P.V.}{2\pi}
			\Int{0}{2\pi}{\lambda'}
			\mu\of{\lambda'} \cot\of{\frac{\lambda-\lambda'}{2}}
			,
		\label{eq:DefHilbert}
		\end{equation}
		or alternatively by its Fourier representation $\Fourier{\sHilbert{\mu};n} = -i\Sign{n} \Fourier{\mu;n}$, where $\Sign{n}$ is the signum function with $\Sign{0} = 0$, and $\Fourier{\cdot;n}$ denotes the Fourier coefficients.

		By definition \eqref{eq:DefWMSDOS} of $\mu\sof{\lambda}$ from $f\sof{E}$, it is apparent that $f\sof{E}$'s singularities, defined by Eq.~\eqref{eq:SpecMeasAssump}, reappear in $\mu\of{\lambda}$.
		Noting that the Hilbert transform can not change the order of the singularity, but may only introduce logarithmic corrections, we find the same singularities in $\sHilbert{\mu}\of{\lambda}$.
		This way, we can justify Eq.~\eqref{eq:AsymResolv}.

		Eq.~\eqref{eq:Plemelj} will be shown in two different ways.
		We first do it by direct manipulation of $u\sof{z}$'s integral definition, then by using Fourier coefficients.

		To begin, observe that the denominator in Eq.~\eqref{eq:DefResolvWMSDOS} can be written as:
		\begin{equation}
			\frac{1}{1-re^{i\lambda}}
			=
			\frac{1}{2} \frac{
				\sbr{1-r}\sbr{1+r}
			}{
				\sbr{1-r}^2 + 4 r \sin^2\of{\frac{\lambda}{2}}
			}
			+ 
			\brr{
				\frac{1}{2}
				+
				i \frac{
					2r\cot\of{\frac{\lambda}{2}}
				}{
					4r 
					+
					\frac{\sbr{1-r}^2}{\sin^2\of{\frac{\lambda}{2}}}
				}
			}
			.
		\label{eq:IntKernelApp}
		\end{equation}
		The first term is a nascent delta function for $r\to1$.
		To see this, consider the following integral:
		\begin{align}
			& \nonumber 
			\frac{1}{2\pi}\Int{\lambda-\pi}{\lambda+\pi}{\lambda'}
			\mu\of{\lambda'}
			\frac{1}{2} \frac{
				\sbr{1-r}\sbr{1+r}
			}{
				\sbr{1-r}^2 + 4 r \sin^2\of{\frac{\lambda-\lambda'}{2}}
			}
			\\ \overset{\epsilon:=\tfrac{1-r}{1+r}}{=} \nonumber &
			\frac{1}{2\pi}\Int{\lambda-\pi}{\lambda+\pi}{\lambda'}
			\mu\of{\lambda'}
			\frac{1}{2} \frac{
				\epsilon
			}{
				\epsilon^2
				+ \sbr{1-\epsilon^2} 
				\sin^2\of{\frac{\lambda-\lambda'}{2}}
			}
			\\ \overset{\lambda'':=\lambda-\lambda'}{=} \nonumber &
			\frac{1}{2\pi}\Int{-\pi}{\pi}{\lambda''}
			\mu\of{\lambda-\lambda''}
			\frac{1}{2} \frac{
				\epsilon
			}{
				\epsilon^2
				+ \sbr{1-\epsilon^2} 
				\sin^2\of{\frac{\lambda''}{2}}
			}
			\\ \overset{t:=\tan\tfrac{\lambda''}{2}}{=} &
			\frac{1}{2\pi}\Int{-\infty}{\infty}{t}
			\mu\of{\lambda-2\arctan\sof{t}}
			\frac{
				\epsilon
			}{
				\epsilon^2
				+ t^2
			}
		\end{align}
		In the first line we shifted the integration boundaries from $[0,2\pi]$ to $[\lambda-\pi,\lambda+\pi]$, which is possible since the integrand is $2\pi$-periodic in $\lambda'$.
		Then we replaced $r$ in a convenient way and changed the integration variable.
		The last variable change is a Weierstrass-substitution which gives $\D\lambda'' = 2 \D t / (1+t^2)$ and $\sin^2\sof{\lambda''/2} = t^2 / (1+t^2)$.
		Now, by taking the limit $r\to1^-$ which corresponds to $\epsilon\to0^+$, one recovers the WMSDOS:
		\begin{equation}
			\frac{1}{4\pi}\Int{0}{2\pi}{\lambda'}
			\frac{
				\mu\sof{\lambda'}
				\sbr{1-r}\sbr{1+r}
			}{
				\sbr{1-r}^2 + 4 r \sin^2\of{\frac{\lambda-\lambda'}{2}}
			}
			\underset{r\to1^-}{\longrightarrow}
			\frac{\mu\of{\lambda}}{2}
		\label{eq:AppPlemelj1}
		\end{equation}

		For the remaining part of Eq.~\eqref{eq:IntKernelApp}, note that $(1 + i \cot\sof{\lambda/2})/2 = (1-e^{i\lambda})^{-1}$.
		As we have already extracted the singular part, a Cauchy principal value remains:
		\begin{align}
			& \nonumber
			\frac{1}{2\pi}\Int{0}{2\pi}{\lambda'}
			\mu\of{\lambda'}
			\brr{
				\frac{1}{2}
				+
				i \frac{
					2r\cot\of{\frac{\lambda-\lambda'}{2}}
				}{
					4r 
					+
					\frac{\sbr{1-r}^2}{\sin^2\of{\frac{\lambda-\lambda'}{2}}}
				}
			}
			\\ \underset{r\to1}{\longrightarrow} & \nonumber
			\frac{P.V.}{4\pi}\Int{0}{2\pi}{\lambda'}
			\mu\of{\lambda'}
			\brr{
				1
				+
				i \cot\of{\tfrac{\lambda-\lambda'}{2}}
			}
			\\ := &
			\frac{1}{2}
			+
			\frac{i}{4\pi}
			P.V. \Int{0}{2\pi}{\lambda'}
			\mu\of{\lambda'} \cot\of{\tfrac{\lambda-\lambda'}{2}} 
			\label{eq:AppPlemelj2}
			.
		\end{align}
		The first term is the normalization of the WMSDOS: $\sInt{0}{2\pi}{\lambda} \mu\of{\lambda} = 2\pi$.
		The convolution with the cotangent is the Hilbert transform on the unit circle, $\Hilbert{\mu}\sof{\lambda}$.
		Putting Eq.~\eqref{eq:AppPlemelj1} and Eq.~\eqref{eq:AppPlemelj2} together, we obtain Eq.~\eqref{eq:Plemelj}:
		\begin{equation}
			u\sof{z}
			\underset{r\to1^-}{\longrightarrow}
			\frac{1}{2} \brr{
				\mu\of{\lambda}
				+
				i \Hilbert{\mu}\of{\lambda}
				+
				1
			}
		\label{eq:AppPlemelj3}
		\end{equation}

		To better see how the Hilbert transform comes into play, it is instructive to re-derive the result in a different way.
		Again, we start from Eq.~\eqref{eq:DefResolvWMSDOS} and expand the geometric series.
		This is done inside and outside of the unit circle:
		\begin{equation}
			\frac{1}{2\pi}\Int{0}{2\pi}{\lambda'} 
			\frac{\mu\of{\lambda'}}{1-ze^{-i\lambda'}}
			= \left\{ \begin{aligned}
				\Sum{n=0}{\infty} z^n \Fourier{\mu;n}, & \quad \sAbs{z} < 1	\\
					- \Sum{n=1}{\infty} z^{-n} \Fourier{\mu;-n}, & \quad \sAbs{z} > 1
			\end{aligned} \right.
		\end{equation}
		Here $\Fourier{\mu;n} := \sbr{2\pi}^{-1} \sInt{0}{2\pi}{\lambda'} e^{-in\lambda'} \mu\sof{\lambda'} = u_n$ are the Fourier coefficients of the WMSDOS measure and also the transition amplitudes.
		Now it is easy to combine the outer and the inner expressions.
		The difference restores the Fourier representation of $\mu\sof{\lambda}$:
		\begin{equation}
			u\sof{r e^{i\lambda}} - u\sof{e^{i\lambda}/r}
			\underset{r\to1^-}{\longrightarrow}
			\Sum{n=-\infty}{\infty} e^{in\lambda} \Fourier{\mu;n}
			=
			\mu\of{\lambda}
			.
		\label{eq:Fatou}
		\end{equation}
		Using the sign function $\Sign{n}$, the sum of the outer and inner limit can be expressed as a Fourier multiplication operator:
		\begin{equation}
			u\sof{r e^{i\lambda}} + u\sof{e^{i\lambda}/r}
			\underset{r\to1^-}{\longrightarrow}
			\Fourier{\mu;0}
			+
			\Sum{n=-\infty}{\infty}e^{in\lambda}
			\Sign{n} \hat{\mu}_n
			.
		\label{eq:Skokhotski}
		\end{equation}
		We used that $\Sign{0}=0$.
		The first term again is the normalization.
		The second term is the Fourier representation of the Hilbert transform times the imaginary unit.
		Its integral representation has been derived above.
		The Fourier representation helps to identify the expression as an Hilbert transform in the first place.
		\begin{equation}
			\Hilbert{\mu}\of{\lambda}
			= \nonumber
			\Sum{n=-\infty}{\infty}e^{in\lambda}
			\brr{ -i \Sign{n} } \hat{\mu}_n
			\label{eq:HilbertApp}
		\end{equation}
		The average of Eq.~\eqref{eq:Skokhotski} and Eq.~\eqref{eq:Fatou} gives Eq.~\eqref{eq:Plemelj}:
		\begin{equation}
			u\sof{re^{i\lambda}}
			\underset{r\to1^-}{\longrightarrow}
			\frac{1}{2}
			\brr{
				\mu\of{\lambda}
				+
				i \Hilbert{\mu}\of{\lambda}
				+
				1
			}
		\label{eq:AppPlemelj4}
		\end{equation}

	\section{The Fourier-Tauber formula}
	\label{app:Tauber}
		In this appendix we explain how to relate the large $n$ asymptotic behavior of a Fourier transform to the singular points of the original function.
		As a starter consider the Fourier transform of an one-sided power term $\sbr{x-x^*}^{\nu-1}\Theta\sof{x-x^*}$, where $\nu>0$ and $\Theta\sof{x}$ is Heaviside's step function.
		We can calculate it by regularizing the integral with an exponential term $e^{-s\sbr{x-x^*}}$, $s > 0$, and taking the limit $s\to0$.
		The result is a power term and a gamma function.
		\begin{align}
			\label{eq:AppAux1}
			& 
			\frac{1}{2\pi} \Int{-\infty}{\infty}{x}
			e^{-inx} \sbr{x-x^*}^{\nu-1}
			\Theta\sof{x-x^*}
			\\ = & \nonumber
			\lim_{s\to0^+}
			\frac{e^{-inx^*}}{2\pi} \Int{x^*}{\infty}{x}
			e^{-\sbr{x-x^*} \sbr{s+in}} \sbr{x-x^*}^{\nu-1}
			\\ = & \nonumber 
			\lim_{s\to0^+}
			\frac{e^{-inx^*}}{2\pi\sbr{s+in}^\nu} \Int{0}{\infty}{x'}
			e^{-x'} x'^{\nu-1}
			=
			\frac{\Gma{\nu}}{2\pi} \frac{e^{-inx^*}}{\sbr{in}^\nu}
			.
		\end{align}
		This result is correct for large absolute values of $n$.
		There is a delta function at $n=0$ that is removed by our regularization.
		In a similar way to before, one can compute the Fourier transform of a power term on the left hand side of $x^*$:
		\begin{align}
			\label{eq:AppAux2}
			\frac{1}{2\pi} \Int{-\infty}{\infty}{x}
			e^{-inx} \sbr{x^*-x}^{\nu-1}
			\Theta\sof{x^*-x}
			=
			\frac{\Gma{\nu}}{2\pi} \frac{e^{-inx^*}}{\sbr{-in}^\nu}
			.
		\end{align}
		Using the last two equations, we can compute the Fourier transform of a function that has a singularity in the $M$-th derivative at $x^*$.
		Consider a function $h\sof{x}$ that behaves like:
		\begin{equation}
			h\sof{x^*\pm\epsilon} 
			\AsymEq
			\Sum{m=0}{M-1} \frac{h^{(m)}\sof{x^*}}{m!} \sbr{\pm\epsilon}^m
			+
			H^\pm \epsilon^{M+\nu-1}
			,
		\label{eq:HAssump}
		\end{equation}
		where $0<\nu<1$ and rest of the notation is like in the main text, see Eq.~\eqref{eq:SpecMeasAssump}.
		Consider the Fourier integral of $h\sof{x}$.
		We split up the integral at $x^*$, use the asymptotic form of $h\sof{x}$ and apply Eqs.~\eqref{eq:AppAux1} and \eqref{eq:AppAux2} to both sides of the integral.
		\begin{align}
			h_n
			= & \nonumber 
			\frac{1}{2\pi} \Int{-\infty}{\infty}{x}
			e^{-inx} h\sof{x}
			\\ \AsymEq & \nonumber 
			\frac{e^{-inx^*}}{2\pi} \left\{ \Int{0}{\infty}{x'} \brr{
				\Sum{m=0}{M-1} \frac{h^{(m)}\sof{x^*}}{m!} x'^m
				+
				H^+ x'^{M+\nu-1}
			} \right.
			\\ & +\nonumber 
			\left. \Int{-\infty}{0}{x'} \brr{
				\Sum{m=0}{M-1} \frac{h^{(m)}\sof{x^*}}{m!} \sbr{-x'}^m
				+
				H^- \sbr{-x'}^{M+\nu-1}
			} \right\}
			\\ = & \nonumber 
			\frac{e^{-inx^*}}{2\pi} \left\{ 
				\Sum{m=0}{M-1} \frac{h^{(m)}\sof{x^*}}{\sbr{in}^{m+1}} 
				+
				\frac{\Gma{M+\nu} H^+}{\sbr{in}^{M+\nu}}
			\right.
			\\ & +\nonumber 
			\left. 
				\Sum{m=0}{M-1} \frac{h^{(m)}\sof{x^*}\sbr{-1}^m}{\sbr{-in}^{m+1}} 
				+
				\frac{\Gma{M+\nu} H^-}{\sbr{-in}^{M+\nu}}
			\right\}
			\\ = & \nonumber 
			\frac{e^{-inx^*}}{2\pi} \left\{ 
				\Sum{m=0}{M-1} \frac{h^{(m)}\sof{x^*}}{\sbr{in}^{m+1}} \brr{ 1 + \frac{\sbr{-1}^m}{\sbr{-1}^{m+1}} }
				+
			\right.
			\\ & +\nonumber 
			\left. 
				\frac{\Gma{M+\nu} H^+}{\sbr{in}^{M+\nu}}
				+
				\frac{\Gma{M+\nu} H^-}{\sbr{-in}^{M+\nu}}
			\right\}
			\\ = & 
			\frac{\Gma{M+\nu}}{2\pi} \frac{e^{-inx^*}}{n^{M+\nu}} \brr{
				\frac{H^+}{i^{M+\nu}}
				+
				\frac{H^-}{\sbr{-i}^{M+\nu}}
			}
			.
			\label{eq:AppAux3}
		\end{align}
		Evidently, the analytic remainder has no influence on the asymptotic result at all.
		The reason is, that contributions from both sides of the singularity cancel.
		Additional splitting the Fourier integral of $h\sof{x}$ at any point where $h\sof{x}$ is smooth will therefore not change the asymptotic result.
		Only the singular points count towards the large $n$ limit.
		
		Consider now that $h\sof{x}$ has multiple singular points $x^*_l$, and admits an expansion like Eq.~\eqref{eq:HAssump} around each of these points.
		The Fourier integral of $h\sof{x}$ is then split up at each $x^*_l$ and the expansion around each point is plugged in.
		Eq.~\eqref{eq:AppAux3} is applied to each part of the integral, leading to a sum of power terms.
		The boundary terms of the integrals are irrelevant, because $h\sof{x}$ is smooth in these points.
		Therefore, we can take each integral's boundaries to infinity.
		\begin{align}
			h_n 
			\AsymEq & \nonumber 
			\frac{1}{2\pi}
			\Sum{l=0}{L-1}
			\Int{x^*_l-\epsilon}{x^*_l+\epsilon}{x} e^{-inx} h\sof{x}
			\\ \AsymEq  & \nonumber 
			\frac{1}{2\pi}
			\Sum{l=0}{L-1}
			\Int{-\infty}{\infty}{x} e^{-inx} [
				\tilde{h}_l\sof{x-x^*_l}  
				+
			\\ & + \nonumber 
				H^+ \sbr{x-x^*_l}^{M+\nu-1} \Theta\sof{x-x^*_l}
				+
			\\ & + \nonumber
				H^- \sbr{x^*_l-x}^{M+\nu-1} \Theta\sof{x^*_l-x}
			]
			\\ = & 
			\frac{\Gma{M+\nu}}{2\pi n^{M+\nu}} \Sum{l=0}{L-1} e^{-inx^*_l} \brr{
				\frac{H^+_l}{i^{M+\nu}}
				+
				\frac{H^-_l}{\sbr{-i}^{M+\nu}}
			}
			.
		\end{align}
		This is the formula used to go from Eq.~\eqref{eq:SpecMeasAssump} to Eq.~\eqref{eq:RAAsym} and also from Eq.~\eqref{eq:AsymIntegrand} to Eq.~\eqref{eq:AsymFDA1}.

	\section{The arrival problem}
	\label{app:Arrival}
		In this appendix we demonstrate that our results hold as well for the problem of first detected arrival, i.e. when $\sKet{\PsiDet} \ne \sKet{\PsiIn}$.
		To do so, it is necessary to introduce another MSDOS, namely the one associated with the {\em pair} $\sKet{\PsiDet}$ and $\sKet{\PsiIn}$.
		Similar to Eq.~\eqref{eq:SpecMeasFinite}, we define
		\begin{equation}
			g\sof{E}
			:=
			\sBAK{\PsiDet}{\delta\sof{E-\Ham}}{\PsiIn}
			.
		\end{equation}
		$g\sof{E}$ can be understood as the joint participation of $\sKet{\PsiDet}$ and $\sKet{\PsiIn}$ in the energy $E$.
		Similar to Eq.~\eqref{eq:DefWMSDOS} we define:
		\begin{equation}
			\nu\sof{\lambda}
			:= 
			\frac{2\pi \hbar}{\tau}
			\Sum{l=-\infty}{\infty} 
			g\sof{\frac{\hbar}{\tau}\sbrr{\lambda+2\pi l}}
			.
		\end{equation}
		While $g\sof{E}$ is the Hamiltonian's spectral measure associated with the detection and the initial state, $\nu\sof{\lambda}$ is the evolution operator's spectral measure associated with these states.
		The transition amplitude from initial to detection state is abbreviated with $v_n := \sBAK{\PsiDet}{\TEO\sof{n\tau}}{\PsiIn}$, its generating function is $v\sof{z} = \sSum{n=0}{\infty} v_n z^n$.

		Repeating the arguments of Appendix~\ref{app:Plemelj} to $v\sof{z}$, we find $v\sof{z}$ close to the unit circle.
		Care has to be taken, as $\nu\sof{\lambda}$ is normalized differently: $(2\pi)^{-1} \sInt{0}{2\pi}{\lambda}\nu\sof{\lambda} = \sBK{\PsiDet}{\PsiIn}$.
		\begin{equation}
			v\sof{re^{i\lambda}}
			\underset{r\to1^-}{\longrightarrow}
			\frac{
				\sBK{\PsiDet}{\PsiIn}
				+
				\nu\of{\lambda}
				+ i \Hilbert{\nu}\of{\lambda} 
			}{2}
			.
		\end{equation}
		The Hilbert transform $\sHilbert{\nu}$ is defined analogously to Eq.~\eqref{eq:DefHilbert}.

		In terms of $v\sof{z}$ and $u\sof{z}$, the generating function of the detection amplitudes is given by [see Eq.~\eqref{eq:QuantumRenewalSol2}]:
		\begin{align}
			\nonumber
			\varphi\sof{re^{i\lambda}}
			= &
			\frac{v\sof{re^{i\lambda}} - \sBK{\PsiDet}{\PsiIn}}{u\sof{re^{i\lambda}}}
			\\ \underset{r\to1^-}{\longrightarrow} &
			\frac{
				\nu\of{\lambda}
				+ i \Hilbert{\nu}\of{\lambda} 
				- \sBK{\PsiDet}{\PsiIn}
			}{
				\mu\of{\lambda}
				+ i \Hilbert{\mu}\of{\lambda} 
				+
				1
			}
		\end{align}

		As we have done before, we require that the singularities of $g\sof{E}$ are identical with the ones of $\rho\sof{E}$.
		Under this assumption, we have the same statement as Eq.~\eqref{eq:AsymResolv}
		\begin{equation}
			v\sof{e^{i\sbr{\Lambda^*_l\pm\epsilon}}}
			\AsymEq
			\tilde{v}_l\sof{\pm\epsilon}
			+
			V^\pm_l
			\epsilon^{\frac{d_S}{2}-1}
			.
		\end{equation}
		The last equation is used together with Eq.~\eqref{eq:AsymResolv} to find the new expansion of $\varphi\sof{e^{i\lambda}}$ around the singular points.
		In this expansion, the following complex constants appear:
		\begin{align}
			\label{eq:DefR}
			R^\pm_l
			:= &
			\frac{1}{B^\pm_l}
			\brr{ 
				V^\pm_l \tilde{u}_l
				- B^\pm_l \br{ \tilde{v}_l - \sBK{\PsiDet}{\PsiIn} }
			}
			\\ = & \nonumber
			\frac{1}{B^\pm_l}
			\brr{ 
				V^\pm_l u\sof{e^{i\Lambda^*_l}}
				- B^\pm_l \br{ v\sof{e^{i\Lambda^*_l}} - \sBK{\PsiDet}{\PsiIn} }
			}
			.
		\end{align}
		Although $u\sof{e^{i\Lambda^*_l}}$ and $v\sof{e^{i\Lambda^*_l}}$ individually may diverge, the particular combination in Eq.~\eqref{eq:DefR} cancels the singularity, whence $R_l^\pm$ is finite.
		We obtain for $\varphi\sof{e^{i\lambda}}$:
		\begin{equation}
			\varphi\sof{e^{i\Lambda^*_l\pm\epsilon}}
			\AsymEq
			\left\{ \begin{aligned}
				\frac{V^\pm}{B^\pm_l}
				-
				R^\pm_l
				\frac{\epsilon^{2 - \frac{d_S}{2}-1}}{B^\pm_l}
				, & \qquad d_S < 2 \\
				\tilde{\varphi}_l
				+
				R^\pm_l
				\frac{B^\pm_l}{\brr{u\sof{e^{i\Lambda^*_l}}}^2}
				\epsilon^{\frac{d_S}{2}-1}
				, & \qquad d_S > 2
			\end{aligned} \right.
			.
		\end{equation}
		The result is the same as in Eq.~\eqref{eq:AsymFDA1}, except for the additional factor $R^\pm_l$.
		This shows, that the probability of first detected arrival will show the same power law decay and exhibit the same oscillations as the probability of first detected return with different amplitudes of the oscillations.
		A discussion of these amplitudes for the one dimensional tight-binding model can be found in Ref.~\cite{Thiel2018-0}.

	\section{Even dimensions}
	\label{app:Even}
			When $d_S$ is an even integer, the return amplitudes still decay like Eq.~\eqref{eq:RAAsym}, but $u\sof{z}$ behaves logarithmically close to the singular points:
			\begin{align}
				u\sof{e^{i\sbr{\Lambda^*_l\pm\epsilon}}}
				\AsymEq & \nonumber
				\tilde{u}_l\sof{\pm\epsilon}
				-
				\frac{C_l}{\Gma{\tfrac{d_S}{2}}}
				\br{e^{\pm i\epsilon}-1}^{\tfrac{d_S}{2}-1}
				\ln\of{1-e^{\pm i\epsilon}}
				\\ \AsymEq &
				\tilde{u}_l\sof{\pm\epsilon}
				-
				\frac{C_l}{\Gma{\tfrac{d_S}{2}}}
				\sbr{\pm i\epsilon}^{\frac{d_S}{2}-1}
				\ln\sof{\mp i \epsilon}
				.
			\label{eq:}
			\end{align}
			For the last line, we used $\sbr{1-e^{i\epsilon}} \AsymEq \sbr{-i\epsilon}$.
			If $d_S > 2$, the expansion of $\varphi\sof{e^{i\lambda}}$ around the singular points looks exactly the same as the expansion of $u\sof{e^{i\lambda}}$, just with $C_l$ replaced by $C_l/u^2\sof{e^{i\Lambda^*_l}}$ [see Eq.~\eqref{eq:AsymFDA1}].
			Therefore, going back to the original space recovers Eq.~\eqref{eq:RAAsym} with different coefficients, which is equal to Eq.~\eqref{eq:AsymFDA2} for large dimensions.

			The case when $d_S =2$ is different.
			It is useful to first look at the case when there is only one critical point $\Lambda^*_0=0$.
			This resembles the classical theory of random walks, where $\varphi_n$ and $u_n$ are real and monotonically decaying.
			The decay of $u_n$ is $u_n \AsymEq C_0 /n$, so that we have:
			\begin{align}
				u\sof{z}
				= &
				\Sum{n=0}{\infty} u_n z^n
				\AsymEq
				C_0 
				\Sum{n=1}{\infty} \frac{z^n}{n}
				=
				C_0 \ln\frac{1}{1-z}
				,
			\end{align}
			as $z\to1^-$.
			The generating function $\varphi\sof{z}$ has the form:
			\begin{equation}
				\varphi\sof{z} 
				\AsymEq 
				1 - \frac{1}{C_0 \ln\tfrac{1}{1-z} }
			\end{equation}
			as $z\to1^-$.
			In this case we may apply classical Tauberian theorems \cite{Feller1971-0} using the fact that $\varphi\sof{z}$ is a slowly varying function.
			The necessary Tauberian theorem states that the {\em sum} $\sSum{m=1}{n}\varphi_m$ behaves like $\varphi\sof{z=1-1/n}$ for large $n$.
			Inverting the relation for $\varphi_n$ yields:
			\begin{align}
				\nonumber
			\varphi_n 
				\AsymEq &
				\frac{1}{2\pi i} 
				\ointctrclockwise\limits_{\Abs{z}=r} \frac{\D z}{z^{n+1}}
				\brr{ 
					1 - \frac{1}{C_0 \ln\sof{\frac{1}{1-z}}}
				}
				\\ \AsymEq &
				\varphi\sof{\tfrac{n-1}{n}} - \varphi\sof{\tfrac{n-2}{n-1}}
				\AsymEq
				\frac{1}{C_0 n \ln^2 n}
				.
			\end{align}
			The second line reveals the logarithmic corrections.
			(We have used $\ln\sof{n-1} \AsymEq \ln n$ and $\ln\sof{n/(n-1)} \AsymEq 1/n$.)
			From the first line, on the other hand, we can learn the behavior of $\varphi_n$, when there are more than one singular points.
			In that case, each singular point contributes a term like above to $\varphi\sof{z}$.
			We can write:
			\begin{equation}
				\varphi\sof{z}
				\AsymEq
				1 - \Sum{l=0}{L-1}
				\frac{1}{C_l} \frac{1}{\ln\tfrac{1}{1-ze^{-i\Lambda^*_l}}}
				.
			\label{eq:}
			\end{equation}
			This expression is plugged into Eq.~\eqref{eq:CauchyInt}.
			In each summand we apply the variable change $z' := z e^{-i\Lambda^*_l}$; this introduces oscillatory factors $e^{-in\Lambda^*_l}$ and restores the prior integral.
			We obtain:
			\begin{equation}
				\varphi_n
				\AsymEq
				\frac{1}{n\ln^2n}
				\Sum{l=0}{L-1}
				\frac{e^{-in\Lambda^*_l}}{C_l}
				.
			\label{eq:AppFDA2D}
			\end{equation}
			This is the asymptotic result we derived and that is shown in Eq.~\eqref{eq:AsymFDP}.
			However, when comparing it to numerical results, the agreement between numerics and our asymptotic theory appears poor.
			The reason is the slow logarithmic convergence.
			To palliate the discrepancy, the logarithmic term in Eq.~\eqref{eq:AppFDA2D} was augmented with a constant term.
			That means, we replaced $\ln^2n$ with $\sbr{\ln n + x}^2$, both expressions are asymptotically equivalent.
			$x$ was fitted such that the envelopes of the numerical and theoretical results agree best in the tails.

\end{document}